\PassOptionsToPackage{pdfpagelabels=false}{hyperref}

\documentclass[fleqn,usenatbib]{mnras}

\usepackage{newtxtext,newtxmath}
\usepackage{xspace}
\usepackage{multirow}

\usepackage[T1]{fontenc}
\usepackage{ae,aecompl}

\usepackage{graphicx}	
\usepackage{amsmath}	
\usepackage{hyperref}

\newcommand{\cmc}{cm$^{-3}$}       

\newcommand{\mum}{\hbox{$\mu$m}}
\newcommand{\lam}{\hbox{$\lambda$}}

\newcommand{\ha}{\hbox{H$\alpha$}}
\newcommand{\hb}{\hbox{H$\beta$}}
\newcommand{\hi}{\hbox{H\,{\sc i}}}
\newcommand{\hii}{\hbox{H\,{\sc ii}}}
\newcommand{\hei}{\hbox{He\,{\sc i}}}

\newcommand{\oiii}{\hbox{[O\,{\sc iii}]}}
\newcommand{\oiv}{\hbox{[O\,{\sc iv}]}}
\newcommand{\neii}{\hbox{[Ne\,{\sc ii}]}}
\newcommand{\neiii}{\hbox{[Ne\,{\sc iii}]}}

\newcommand{\nii}{\hbox{[N\,{\sc ii}]}}

\newcommand{\oiiiw}{\hbox{[O\,{\sc iii}]\,\lam88\mum}}
\newcommand{\oiiiwv}{\hbox{[O\,{\sc iii}]\,\lam5007\AA}}

\newcommand{\oivw}{\hbox{[O\,{\sc iv}]\,\lam25.9\mum}}
\newcommand{\neiiw}{\hbox{[Ne\,{\sc ii}]\,\lam12.8\mum}}
\newcommand{\neiiiw}{\hbox{[Ne\,{\sc iii}]\,\lam15.6\mum}}

\newcommand{\niiw}{\hbox{[N\,{\sc ii}]\,\lam6583\AA}}

\newcommand{\Msol}{\hbox{M$_\odot$}}    

\newcommand{\zstar}{\hbox{$Z_{\ast}$}}   

\newcommand{\mup}{\hbox{${\rm M}_{u}$}}
\newcommand{\mlow}{\hbox{${\rm M}_{l}$}}
\newcommand{\logu}{\hbox{${\rm log}\,(U)$}}

\newcommand{\cb}{C\&B\xspace}

\newcommand{\guion}{\multicolumn{1}{c}{--}}

\title[Optical and IR diagnostic diagrams]{On the origin of optical and IR emission lines in star forming galaxies}

\author[M. Mart\'inez-Paredes et al.]{Mariela Mart\'inez-Paredes$^{1}$\thanks{E-mail: mariellauriga@gmail.com},  Gustavo Bruzual$^{2}$\thanks{E-mail: g.bruzual@irya.unam.mx}, Christophe Morisset$^{3}$, Minsun Kim$^{1}$, Marcio Mel\'endez$^{4}$ 
\newauthor
and Luc Binette$^{3}$
\\
$^{1}${Korea Astronomy and Space Science Institute 776, Daedeokdae-ro, Yuseong-gu, Daejeon, Republic of Korea (34055)}\\
$^{2}$Instituto de Radioastronom\'{\i}a y Astrof\'{\i}sica, Universidad Nacional Aut\'onoma de M\'exico, Morelia, 58089 Michoac\'an, M\'exico\\
$^{3}$Instituto de Astronom\'ia, Universidad Nacional Aut\'onoma de M\'exico, Apdo. Postal 70264, 04510 CdMx, M\'exico\\
$^{4}$Space Telescope Science Institute, Baltimore, MD 21218, USA\\
}
\date{Accepted 2023 August 10. Received Received 2023 August 9; in original form 2023 May 5}

\pubyear{2022}

\begin{document}
\label{firstpage}
\pagerange{\pageref{firstpage}--\pageref{lastpage}}
\maketitle

\begin{abstract}
Combining the {\sc Cloudy} photoionization code with updated stellar population synthesis results, we simultaneously model the MIR $\neiii/\neii$ vs. $\oiv/\neiii$, the MIR-FIR $\neiii/\neii$ vs. $\oiv/\oiii$ and the classical BPT diagnostic diagrams. We focus on the properties of optically classified \hii\,galaxies that lie in the normal star forming zone in the MIR diagnostic diagram.
We find that a small fraction of our models lie in this zone, but most of them correspond to the lowest explored metallicity, \zstar\,=\,0.0002, at age $\sim1$ Gyr. This value of \zstar\,is, by far, lower than the values derived for these galaxies from optical emission lines, suggesting that the far-UV emission produced by post-AGB stars (a.k.a. HOLMES, hot low-mass evolved stars) is NOT the source of ionization. 
Instead, shock models can easily reproduce this part of the MIR diagram. We suggest that it is likely that some of these galaxies have been misclassified and that in them, shocks, produced by a weak AGN-outflow, could be an important source of ionizaton.
Using a subset of our models, we derive a new demarcation line for the maximal contribution of retired galaxies in the BPT diagram. This demarcation line allows for a larger contamination from the neighbouring AGN-dominated region.
Considering the importance of disentangling the different ionising mechanisms in weak or deeply obscured systems, new observational efforts to classify galaxies both in the optical and IR are required to better constrain this kind of models and understand their evolutionary paths.
\end{abstract}




\section{Introduction}

The optical-line diagnostic diagram \citep[BPT diagram][]{BPT81, Veilleux87},
based on the $\oiiiwv/\hb$ and $\niiw/\ha$ emission line ratios,
has been a successful tool for distinguishing the thermal emission resulting from star formation from the non-thermal emission produced in the central engine of active galactic nuclei \citep[AGNs, e.g.,][]{Kauffmann03,Groves04,Kewley06,Stasinska06,Farrah07,Levesque10,Richardson16,Zhang20}. However, the BPT diagram, based on the least extinguished optical emission lines, fails to distinguish these contributions in optically obscured systems (e.g., dusty galaxies), where starburst galaxies (SB) and AGNs coexist \citep[e.g.,][]{Satyapal08, Goulding09, Melendez14}. The BPT diagram has also limitations identifying retired galaxies, in which star formation stopped several Gyr ago and excitation is due to the presence of hot post Asymptotic Giant Branch (pAGB) stars, also known as HOLMES \citep[hot low-mass evolved stars,][]{Stasinska22, Stasinska15, Flores11, Stasinska08}. Mid-infrared (MIR) wavelengths offer a window in which the effects of extinction are minimal. The high resolution {\it Spitzer} spectra have been widely used to build MIR diagnostic diagrams that allow to differentiate the relative contributions from AGN and star forming (SF) regions \citep[e.g.,][]{Genzel98, Sturm02,Peeters04, Sturm06, Weedman06b, Dale06, Brandl06, Spoon07, Armus07, Hunter07, Farrah07, Melendez08a, Hao09}. 

\begin{figure}
  \centering
  \includegraphics[width=\columnwidth]{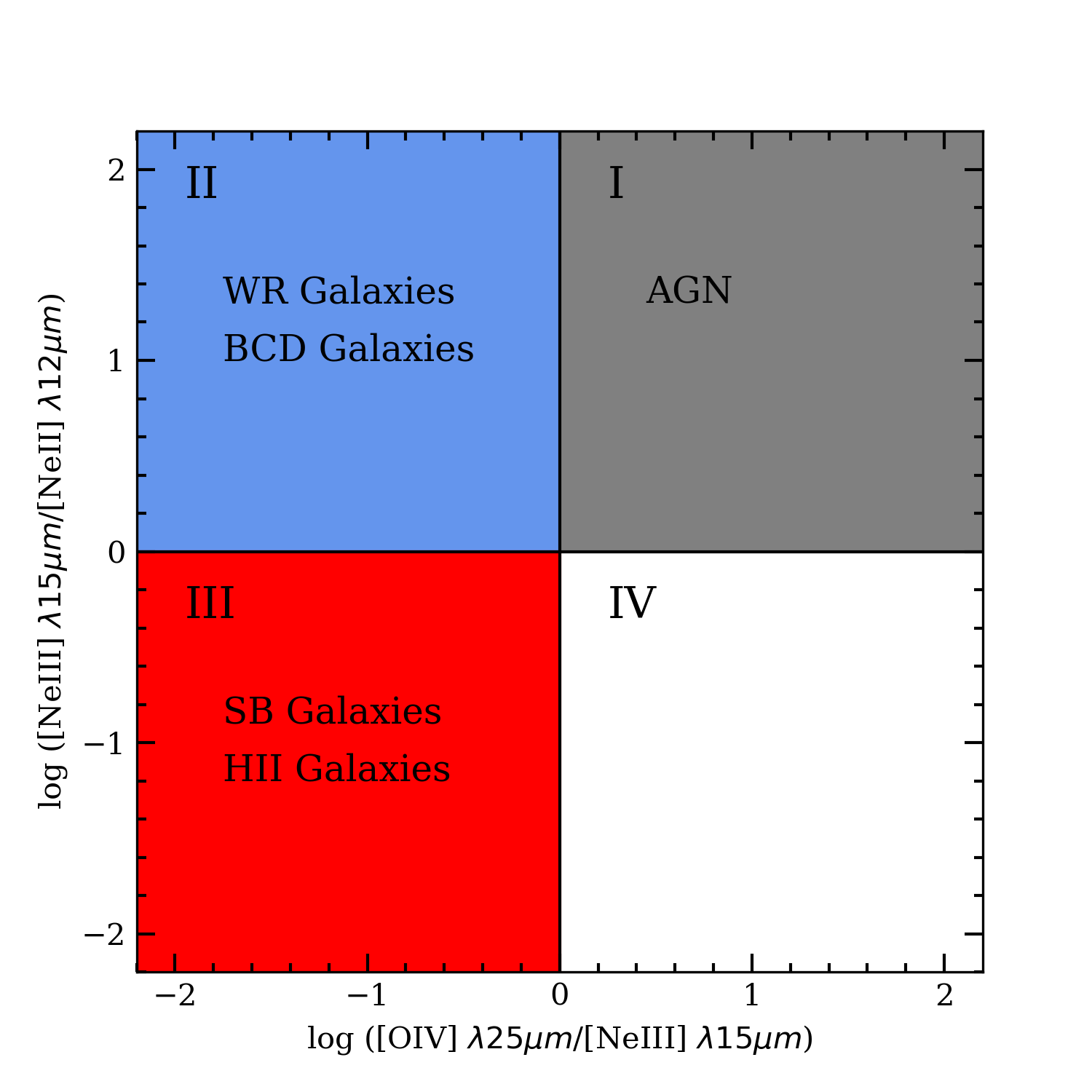}
  \caption{Schematic MIR 
  diagnostic diagram defined by \citet[][their figure 9]{Weaver10}. The boundaries of the different zones are defined in Table~\ref{tab:mirdd}.
  Wolf-Rayet (WR) and BCD galaxies fall preferentially in quadrant II, 
  SB and HII galaxies in quadrant III, and AGN in quadrant I. The low ionization narrow emission regions (LINERs) extend over the AGN zone and quadrant III. ULIRGs fall mostly in quadrant III and quasars in quadrant I. Quadrant IV is mostly void of objects.}
  \label{fig:DD_def}
\end{figure}

\citet[][]{Weaver10} used a hard X-ray (14-195 keV) selected sample of local ($z$\,$\sim$\,$0.025$) type\,1\,and\,2 Seyfert galaxies with high resolution {\it Spitzer} spectra, to investigate how the
$\neiiiw/\neiiw$ and $\oivw/\neiiiw$ 
line ratios compare with the ratios measured in a sample of Blue Compact Dwarf \citep[BCD,][]{Hao09}, \hii\ \citep[also named normal-SF,][]{Goulding09} and 
SB \citep[][]{Bernard-Salas09} galaxies. Three zones (quadrants) were identified in this MIR diagnostic diagram, listed in Table~\ref{tab:mirdd} and shown schematically in Fig.~\ref{fig:DD_def} \citep[see][their figure 9]{Weaver10}.

\begin{table}
\caption{\label{tab:mirdd}MIR diagnostic diagram \citep{Weaver10}.}
\centering
\begin{tabular}{ccc}
\hline
Quadrant & $\log \frac{\oivw}{\neiiiw}$ & $\log \frac{\oivw}{\neiiiw}$ \\
\hline
I   & $\gtrsim 0$  & $\gtrsim 0$  \\
II  & $\lesssim 0$ & $\gtrsim 0$  \\
III & $\lesssim 0$ & $\lesssim 0$ \\
IV  & $\gtrsim 0$  & $\lesssim 0$ \\
\hline
\end{tabular}
\end{table}
\begin{table*}
\begin{center}
 \caption{\label{tab:observations}Observed nuclear MIR, FIR and optical emission line ratios of selected \hii\ and SB galaxies, and AGN}
 \begin{tabular}{llrrrrr}
 \hline
      &      & $\lambda$15.6$\mu$m/$\lambda$12.8$\mu$m  & $\lambda$25.9$\mu$m/$\lambda$15.6$\mu$m & $\lambda$25.9$\mu$m/$\lambda$88$\mu$m & $\lambda$5007\AA/$\lambda$4861\AA  & $\lambda$6584\AA/$\lambda$6563\AA \\
 Type        & Name        & log\,([NeIII]/[NeII]) & log\,([OIV]/[NeIII]) & log\,([OIV]/[OIII]) & log\,([OIII]/H$_{\beta}$) & log\,([NII]/H$_{\alpha}$) \\
 \hline
AGN$^a$            &  NGC 1097   & $-1.072\pm0.004$  &   $-0.983\pm0.019$  & \guion &  $0.444\pm0.835$   &   $0.191\pm0.182$  \\
                   &  NGC 1291   &  $0.139\pm0.172$  &   $-0.706\pm0.037$  & \guion &  $0.629\pm2.740$   &   $0.222\pm0.460$  \\
                   &  NGC 1566   & $-0.262\pm0.034$  &   $-0.187\pm0.086$  & \guion &  $0.626\pm0.627$   &  $-0.065\pm0.094$  \\
                   &  NGC 2841   &  $0.072\pm0.115$  &   $-0.937\pm0.022$  & \guion &  $0.324\pm0.769$   &   $0.175\pm0.448$  \\
                   &  NGC 3031   & $-0.103\pm0.027$  &   $-0.725\pm0.037$  & \guion &  $0.499\pm1.626$   &   $0.192\pm0.460$  \\
                   &  NGC 3621   & $-0.448\pm0.026$  &   $-0.091\pm0.050$  & \guion &  $1.039\pm5.041$   &   $0.031\pm0.129$  \\
                   &  NGC 4450   & $-0.220\pm0.157$  &   $-0.523\pm0.100$  & \guion &  $0.508\pm0.944$   &   $0.340\pm0.438$  \\
                   &  NGC 4579   & $-0.318\pm0.021$  &   $-0.630\pm0.010$  & \guion &  $0.638\pm0.544$   &   $0.322\pm0.236$  \\
                   &  NGC 4594   &  $0.064\pm0.077$  &   $-0.863\pm0.015$  & \guion &  $0.420\pm1.183$   &   $0.320\pm0.576$  \\
                   &  NGC 4725   &  $0.069\pm0.268$  &   $-0.236\pm0.119$  & \guion &  $0.975\pm7.409$   &   $0.145\pm0.828$  \\
                   &  NGC 4826   & $-0.662\pm0.006$  &   $-0.728\pm0.004$  & \guion &  $0.013\pm0.285$   &   $0.083\pm0.132$  \\
                   &  NGC 5033   & $-0.519\pm0.012$  &   $-0.293\pm0.040$  & \guion &  $0.628\pm1.939$   &   $0.271\pm0.345$  \\
                   &  NGC 5194   & $-0.277\pm0.011$  &   $-0.346\pm0.018$  & \guion &  $0.863\pm1.268$   &   $0.432\pm0.199$  \\
                   &  NGC 5866   & $-0.200\pm0.084$  &   $-0.720\pm0.038$  & \guion &  $0.499\pm2.483$   &   $0.250\pm0.657$  \\
\hline
SB galaxies$^a$      &  NGC 1705   &  $0.585\pm0.977$   &  $-0.703\pm0.035$  & \guion &  $0.240\pm0.176$   &  $-1.127\pm0.039$  \\
                   &  NGC 2915   &  $0.533\pm0.380$   &  $-1.387\pm0.012$  & \guion &  $0.488\pm0.296$   &  $-1.308\pm0.022$  \\
                   &  Mrk 33     & $-0.123\pm0.010$   &  $-1.701\pm0.006$  & \guion &  $0.254\pm0.027$   &  $-0.733\pm0.004$  \\
                   &  NGC 3773   & $-0.054\pm0.034$   &  $-1.640\pm0.043$  & \guion &  $0.173\pm0.028$   &  $-0.825\pm0.005$  \\
\hline
\hii\ galaxies$^a$ & NGC 3351    &  $-1.073\pm0.005$  &  $-0.861\pm0.066$  & \guion &  $-0.802\pm0.047$  &  $-0.332\pm0.038$  \\
                   & NGC 4254    &  $-0.941\pm0.007$  &  $-0.383\pm0.040$  & \guion &  $-0.002\pm0.233$  &  $-0.406\pm0.037$  \\
                   & NGC 4569    &  $-0.354\pm0.016$  &  $-0.747\pm0.022$  & \guion &  $-0.023\pm0.115$  &  $-0.021\pm0.058$  \\
                   & NGC 5713    &  $-0.875\pm0.003$  &  $-0.785\pm0.037$  & \guion &  $-0.339\pm0.048$  &  $-0.346\pm0.014$  \\
\hline
\hii\ galaxies$^b$ & NGC253      &  $-1.141\pm0.004$  &  $-0.121\pm0.136$  & $-0.332\pm0.081$  & \guion & \guion \\
                   & NGC 1614    &  $-0.595\pm0.010$  &  $-0.863\pm0.014$  & $-1.227\pm0.006$  & \guion & \guion \\
                   & NGC 2146    &  $-0.836\pm0.004$  &  $-0.674\pm0.049$  & $-1.770\pm0.004$  & \guion & \guion \\
                   & M 82        &  $-0.796\pm0.015$  &  $-0.249\pm0.083$  & $-1.640\pm0.003$  & \guion & \guion \\
                   & NGC 3184    &  $-0.930\pm0.014$  &  $-0.127\pm0.161$  & $-0.377\pm0.248$  & \guion & \guion \\
                   & NGC 3256    &  $-0.902\pm0.003$  &  $-0.722\pm0.047$  & $-1.192\pm0.016$  & \guion & \guion \\
                   & M95         &  $-1.043\pm0.007$  &  $0.440\pm1.870$   & $-0.301\pm0.341$  & \guion & \guion \\
                   & NGC 4536    &  $-0.764\pm0.003$  &  $-0.551\pm0.041$  & $-1.773\pm0.002$  & \guion & \guion \\
                   & M 83        &  $-1.235\pm0.003$  &  $-0.707\pm0.037$  & $-1.238\pm0.011$  & \guion & \guion \\
                   & NGC 6946    &  $-1.009\pm0.003$  &  $0.026\pm0.183$   & $-0.479\pm0.061$  & \guion & \guion \\
\hline
AGN$^{c}$ & NGC 6552$^{~d1}$ &$0.059\pm0.024$  & $0.073\pm0.014$ & \guion & \guion & \guion\\
          & NGC 6552$^{~d2}$ &$0.004\pm0.021$  & $-0.137\pm0.011$& \guion & \guion & \guion \\
          & NGC 6552$^{~d3}$ &$0.023\pm0.024$  & $0.010\pm0.011$& \guion & \guion & \guion  \\
          \\
          & NGC 7319$^{~e1}$ & $0.321\pm0.015$ & $0.402\pm0.007$& \guion & \guion & \guion \\
          & NGC 7319$^{~e2}$ & $0.066\pm0.009$ & $-0.014\pm0.008$& \guion & \guion & \guion \\
          & NGC 7319$^{~e3}$ & $0.520\pm0.033$ & $0.243\pm0.013$& \guion & \guion & \guion \\
 \hline
\multicolumn{7}{l}{$^a$From SINGS \citep{Dale06,Dale09}. $^b$From {\it Herschell}\,/\,SPIRE \citep{Fernandez16}. $^c$From {\it JWST}/MIRI: $^{d1}$Nuclear,}\\ 
\multicolumn{7}{l}{$^{d2}$Circumnuclear, $^{d3}$Central \citep{Alvarez_Marquez22}; $^{e1}$AGN, $^{e2}$N2, $^{e3}$S2 \citep{Pereira-Santaella22}.}\\
\end{tabular}
\end{center}
\end{table*}
\begin{table*}
\begin{center}
 \caption{\label{tab:imf}\cb\ model parameters}
 \begin{tabular}{lccccc}
 \hline
\multirow{2}{*}{IMF}& \multirow{2}{*}{$\alpha$} & Mass range          & \mlow   & \mup         & $\zstar$\\
                    &                           & [\Msol]             & [\Msol] & [\Msol]      & $\log\,(\zstar)$\\
 \hline
 \citet{Kroupa01}   &         +2.30             & $0.5 \leq m \leq \mup$  & 0.1 & 100, 300     & 0.0002, 0.002, 0.008, 0.014, 0.017, 0.03, 0.06 \\
                    &         +1.30             & $\mlow \leq m  <   0.5$ &     &              & -3.7, -2.7. -2.1, -1.85, -1.77, -1.52, -1.22 \\
 \hline
Top-heavy ($x030$)  &         +1.30             & $\mlow \leq m \leq \mup$  & 0.1  & 100, 300  & '' \\
 \hline
\multicolumn{6}{l}{Note.- Column 1 lists the IMF, defined as $\Phi(m) = dN/dm = Cm^{-\alpha} = Cm^{-(1+x)}$, where $C$ is a normalization constant. Columns 2 to 5}\\
\multicolumn{6}{l}{give the IMF slope, mass range, and lower and upper mass limits, respectively. For the \citet{Salpeter55} IMF, $\alpha$\,=\,2.35. Column 5 lists the}\\
\multicolumn{6}{l}{stellar metallicities used in the \cb\ models.}\\
 \end{tabular}\\

\end{center}
\end{table*}

\begin{table*}
\begin{center}
 \caption{\label{tab:pp}Physical parameters used in the photoionization models}
 \begin{tabular}{lll}
 \hline
 Parameter.         & Values explored & Description\\
 \hline
 SED type & 'SSP', 'CSF' & Single Stellar Population and Constant Star Formation \\
 log age [yr] & 6, 6.18, 6.3, 6.4, 6.48, 6.54, 6.6, 6.65, 6.7, 6.74, 6.78, 6.81, & in SSP case \\

            & 6.85, 6.88, 6.9, 6.93, 6.95, 7, 7.48, 8, 8.18, 8.3, 8.4, 8.48, 8.54, & ... \\
            & 8.6, 8.65, 8.7, 8.74, 8.78, 8.81, 8.85, 8.88, 8.9, 8.93, 8.95, 8.98, & ... \\
            & 9, 9.48, 9.7 and 10. & \\

log age [yr] & 6, 6.18, 6.3, 6.4, 6.48, 6.54, 6.6, 6.65, 6.7, 6.74, 6.78, 6.81, & in CSF case\\
& 6.85, 6.88, 6.9, 6.95, 7, 7.48, 8, 8.48, 9, 9.48, 9.7 and 10. & ...\\
 \logu         & -1.5, -2, -2.5, -3, -3.5, -4 & ionization parameter\\
 log\,($n_H$) [\cmc]      & 1, 2, 3, 4     & Hydrogen density \\ 
 log\,$(\rm O/H)_{total}$ & -5.06, -4.06, -3.45, -3.20, -3.12, -2.86, -2.58 &  total Oxygen abundance \\
 $\rm C/O$             & -1, -0.36, 0.15 &\\
 $\Delta{\rm N/O}$     & -0.25, 0, 0.25 & deviation from N/O to O/H ratio \citep{Gutkin16}   \\
 $\xi$ & 0.36 &  dust-to-metal mass ratio \\
 $fr$ & 1.0 & form factor \\
\hline
\end{tabular}
\end{center}
\end{table*}

Using the {\sc  Cloudy\footnote{\url{https://gitlab.nublado.org/cloudy/cloudy/-/wikis/home}}} photoionization code \citep[]{Ferland17} and the stellar libraries from {\sc Starburst99\footnote{\url{https://www.stsci.edu/science/starburst99/docs/default.htm}}} \citep[][]{Leitherer99}, \citet{Melendez14} produced an extensive set of photoionization models where they successfully modelled the AGN and the BCD galaxy zones, but failed modelling the SB and \hii\ galaxy zone. They tested a two-zone SB model to see if they could understand the normal-SF region (quadrant III). However, this resulted in a highly degenerated solution, due to the large number of stellar and physical parameters involved. \citet[][]{Fernandez16} carried out a similar approach, but in addition to the MIR $\oivw/\neiiiw$ ratio, they included the far-IR (FIR) $\oivw/\oiiiw$ emission line ratio proposed by \citet[][]{Spinoglio15}, assuming a single zone model. As in \citet{Melendez14}, \citet[][]{Fernandez16} were unable to model the normal-SF zone. Both groups argued that accurate modelling of the normal-SF zone requires a better treatment of the stellar atmospheres above 54.94 eV.

The \oivw\ line, with ionization energy\footnote{National Institute of Standards and Technology (NIST) Atomic Spectra Database.} $E_{IP}=77.41$ eV, is among the species requiring the highest ionization energy that is observable with the {\it Spitzer} Infra Red Spectrograph (IRS) in the spectrum of galaxies and AGNs \citep[][]{Genzel98, Lutz98}. This line tracks recent star formation activity, when the intrinsic luminosity of the AGN can be up to 20 times lower than the host galaxy \citep[][]{Pereira-Santaella10}, and is a good indicator of AGN activity in Seyfert galaxies \citep[][]{Genzel98, Melendez08a, Pereira-Santaella10, Weaver10}. 
On the other hand, the intermediate-ionization \neiiiw\ (40.96 eV) and \oiiiw\ (35.12 eV) lines are present in the {\it Spitzer} spectrum of composite systems where SB and AGN coexist, while the
low-ionization \neii\ (21.57 eV) line tracks recent star formation activity \citep[][]{Ho07}.
In the \citet[]{Weaver10} MIR diagnostic diagram, some \hii\,galaxies go undetected in \oivw\ and, when detected, there is sometimes a mixed-up optical classification, with some galaxies being classified as Seyfert or LINER.

In this Paper we use the {\sc Cloudy} photoionization code \citep[][c17.03]{Ferland17} and the revised version of the \citet[][]{Bruzual03} stellar population synthesis models introduced in \citet[][hereafter \cb models]{plat2019} to analyze a small sample of galaxies 
presenting optical and MIR emission lines, aiming at discovering:
{\it (i)} the star forming and physical properties characterizing the galaxy population in the normal-SF zone of the MIR diagnostic diagram, 
{\it (ii)} the maximum contribution to the ionizing radiation from HOLMES allowed in this zone, and 
{\it (iii)} the position in the BPT diagram of galaxies that in the MIR diagnostic diagram lie in the normal-SF and SB zones.
The paper is organized as follows. In Sec.\,2 we select our sample. In Sec.\,3 we describe the stellar models and the {\sc  Cloudy} input parameters. We present the analysis of our results for the MIR in Sec.\,4 and for the optical in Sec.\,5, where we define a new HOLMES demarcation line. The summary and conclusions are presented in Sec.~\ref{sec:conclusions}.

\section{MIR and Optical Data}\label{sec:sample}
To compare with our photoionization model predictions, we collect from the available literature the optical (\oiiiwv, \niiw, \ha, and \hb) and MIR (\neiiiw, \neiiw, and \oivw) emission line measurements, corresponding to the same nuclear region, for AGN, SB, and \hii\,galaxies. Our final sample is listed in Table~\ref{tab:observations}. From the {\it Spitzer} Infrared Nearby Galaxy Survey \citep[SINGS,][]{Dale06, Dale09, Moustakas10}, we select data in both spectral ranges for 14 AGN, 4 SB, and 4 \hii\,galaxies. We add a sample of \hii\,galaxies with available nuclear measurements of the FIR \oiiiw\ emission line from {\it Herschel}\,/\,SPIRE \citep[][]{Fernandez16}. 

We include MIR data for two AGNs recently observed with the Mid-Infrared Instrument (MIRI) on the {\it James Webb Space Telescope} \citep[JWST,][]{Alvarez_Marquez22, Pereira-Santaella22}. NGC~6552 observed during commissioning (Program I.D \#1039, P.I. D. Dicken) and
NGC~7319, observed as part of the Early Release Observations \citep[ERO; Program ID \#2732;][]{Pontoppidan22}. For NGC~6552, \cite{Alvarez_Marquez22} report nuclear, circumnuclear (annulus of width $\sim$\,0.33 kpc), and central (aperture $\sim$\,0.88 kpc, 1.6 arcsec) emission. For NGC~7319, \cite{Pereira-Santaella22} report emission from the nuclear region (denoted AGN) and two asymmetric radio hotspots: N2 and S2, located at 0.43 and 1.5 kpc from the nucleus, respectively.

Additionally, we select 404 AGN with clear detection of nuclear emission in the four optical emission lines from the Burst Alert Telescope on {\it Swift} \citep[BAT,][]{Koss17}. $\sim17$ per cent of this sample also shows nuclear emission in the four MIR emission lines \citep[][]{Weaver10}.

Likewise, we select a sample of 767,738 optical emission line galaxies from SDSS DR8 \citep[][]{Brinchmann04,Tremonti04}, observed with signal-to-noise ratio $>$\,$3\sigma$, and 80 SF galaxies\footnote{These galaxies are part of the {\it Spitzer}-SDSS-{\it GALEX} spectroscopy Survey and the {\it Spitzer} SDSS Spectroscopy Survey (SSGSS, S5; PI: D. Schiminovich).} with available $\neiiiw/\neiiw$ emission line ratio, but undetected \oivw\ emission \citep[][]{LaMassa12}. For conciseness, these objects are not listed in Table~\ref{tab:observations}.
 
\section{models}\label{sec:models}
\subsection{Stellar populations}

The stellar ingredients used in the \cb models are explained in detail in \citet[][appendix A, tables 8 to 12]{Sanchez22}.
The \cb models include advances in the stellar evolution theory from the last decade. The models cover in detail the evolution of O and B stars, including the Wolf-Rayet (WR) phase, and treat HOLMES (pAGB stars) following \citet{Miller16}. Each \cb model contains 220 spectra covering the age range from 0 to 14 Gyr. In a simple stellar population (SSP) the most massive (O, B and Wolf-Rayet) stars last a few Myr, depending on their mass. 
During this time, the strong UV continuum emitted by these young hot stars is responsible for the ionization and excitation of the surrounding medium. 
At age between 70 and 500 Myr, depending on metallicity, HOLMES appear and become responsible for the far-UV emission, being the only source of photoionization \citep[][]{Stasinska08}.
The SSP models were scaled to a typical cluster mas of $1\times 10^{6}$\,\Msol.
Besides the instantaneous burst models (SSPs), we explore models with a continuous star formation rate (SFR) at a constant value of $1\,\Msol yr^{-1}$ (hereafter CSF models).

To test different stellar mass distributions, we use models computed from assuming a \citet{Kroupa01} and a Top-heavy initial mass function (IMF), with parameters listed in Table~\ref{tab:imf}. Hereafter, the Top-heavy model will be denoted as $x030$ model to reflect its IMF slope.
For both IMFs we tested models with upper mass limit of star formation, $M_u$\,=\,100 and 300 $\Msol$. In all cases, the lower mass limit is $M_l$\,=\,0.1 $\Msol$. To explore a large set of spectral energy distributions (SEDs), we select \cb models for the seven values of metallicities $\zstar$ listed in Table~\ref{tab:imf}.

\begin{figure*}
\centering
  \includegraphics[width=\textwidth]{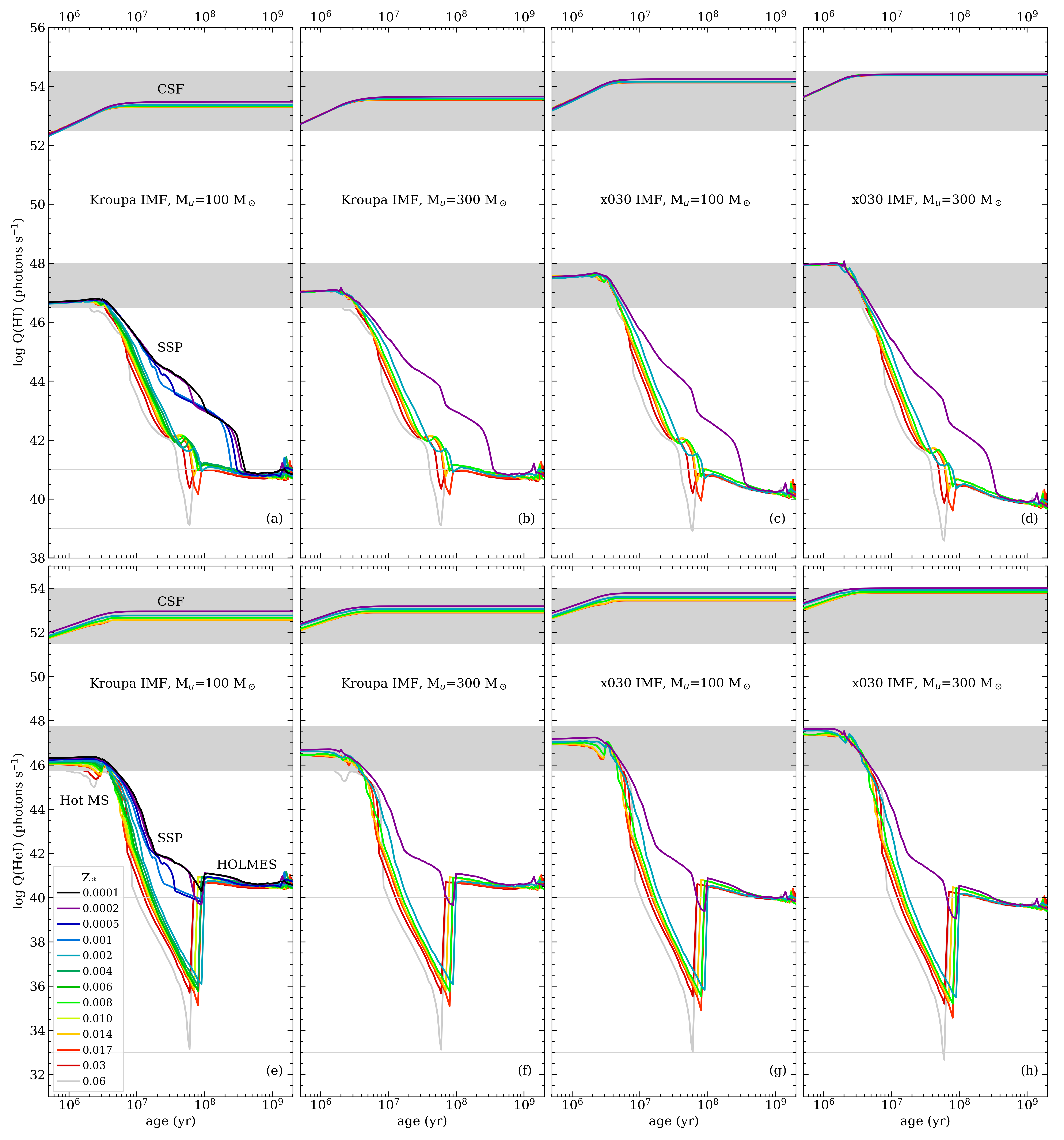}
  \caption{Production rate of \hi\ ({\it top panels}) and  \hei\ ({\it bottom panels}) ionizing photons in the \cb\ stellar population models used in this work (see Table~\ref{tab:imf}): 
  {\it (a,e)} Kroupa IMF, \mup\,=\,100\,\Msol,
  {\it (b,f)} Kroupa IMF, \mup\,=\,300\,\Msol,
  {\it (c,g)} x030 IMF, \mup\,=\,100\,\Msol, and
  {\it (d,h)} x030 IMF, \mup\,=\,300\,\Msol.
  Models are colour coded according to \zstar\ as indicated in the legend in panel\,{\it (e)} legend. In each panel, the lower set of lines corresponds to the SSP models and the upper set to the CSF models.
  For the SSP models in panels {\it (a,e)} we indicate the age range at which the hot MS stars and the HOLMES are responsible for the ionizing radiation. The {\it grey horizontal lines and bands} are drawn to guide the eye when comparing the different models. 
  For the SSP models, $Q$ corresponds to a stellar population of mass\,=\,1\,\Msol. For the CSF models the mass of the population increases continuously at a rate of 1\,\Msol\ yr$^{-1}$.
  For illustration purposes, in panels ({\it a,e}) we include 6 extra SSP and CSF models, not listed in Table~\ref{tab:imf}, corresponding to $\zstar$\,=\,0.0001, 0.0005, 0.001, 0.004, 0.006, 0.01.
  }
 \label{fig:ionphot}
\end{figure*}

\begin{figure}
\centering
\includegraphics[width=0.4\textwidth]{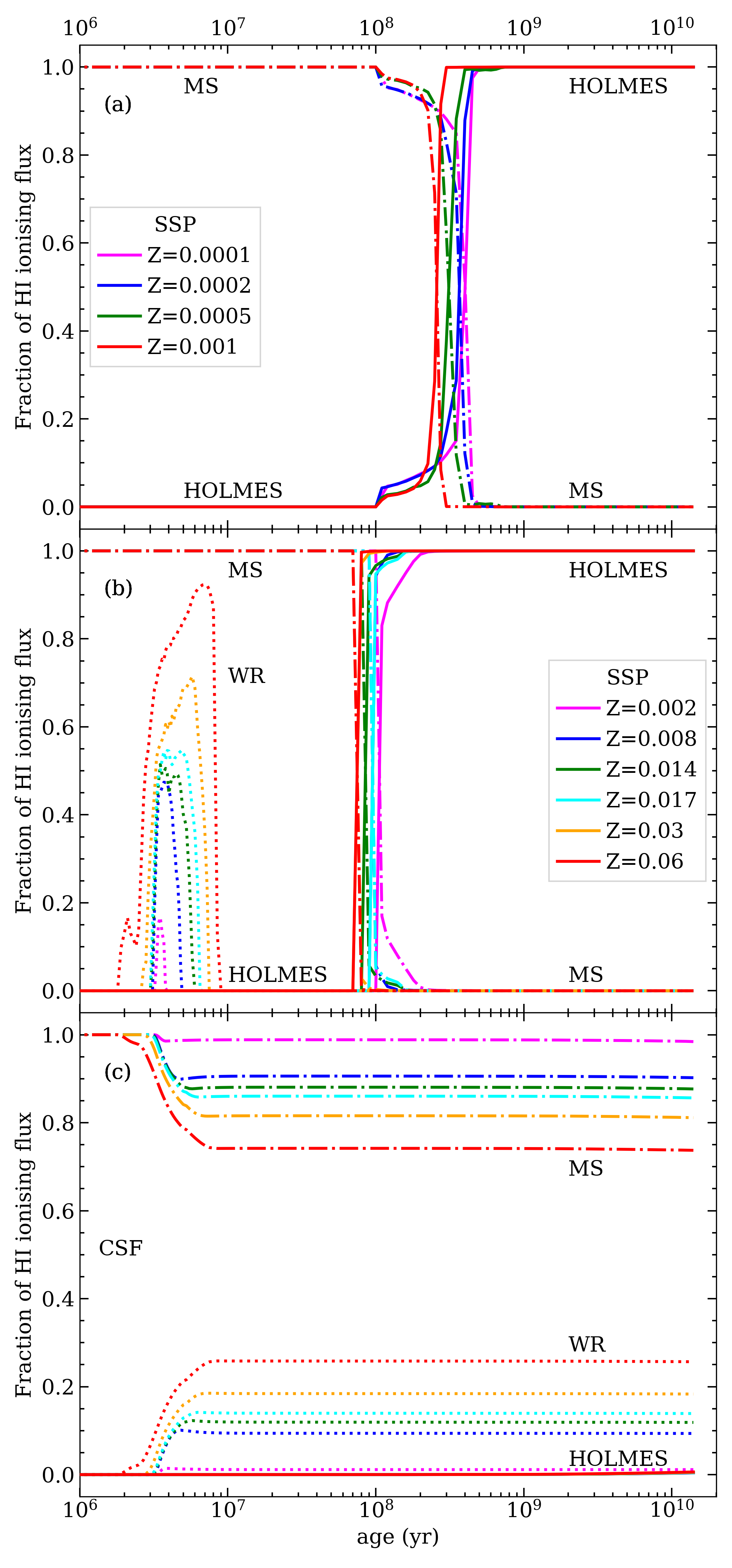}
\caption{Fraction of \hi\ ionising flux emitted by MS, WR and HOLMES stars vs. age ({\it dot-dashed, dotted} and {\it solid} lines, respectively.
Panel {\it (a)} shows the 4 lowest $\zstar$ \cb\ SSP  models. At these metallicities, HOLMES appear at age from 200 to 500 Myr and no WR stars are expected in the stellar population. Only the \zstar\,=\,0.0002 model is used in this investigation.
Panel {\it (b)} corresponds to the other 6 \zstar\ SSP models in Table~\ref{tab:imf}. For these metallicities the appearance of HOLMES occurs at about 70 to 100 Myr, and the contribution from WR stars is clearly seen at age ranging from 3 to 8 Myr, depending on \zstar. The transition from the MS to the HOLMES dominated regimes is clearly seen in panels {\it (a,b)}. Panel {\it (c)} shows the contribution of these stellar groups to the \hi\ ionizing flux of the CSF models for the same \zstar\ values of panel {\it (b)}. In the CSF models the contribution of WR stars can reach 25\%, whereas the contribution from HOLMES is negligible at all ages. Models in panel {\it (c)} are colour coded according to \zstar\ as indicated in the legend in panel {\it (b)}. In all panels we show models that assume the Kroupa IMF with $M_u$\,=\,100\,\Msol.}
\label{fig:ionflux}
\end{figure}

In Figs.~\ref{fig:ionphot} and \ref{fig:ionflux} we show the ionization properties of the \cb\ models. In SSP models, depending on the IMF, the rate of production of ionising photons in the HOLMES era is up to eight orders of magnitude lower than at early ages when the hot main sequence (MS) stars are the dominant source. The transition between these two regimes happens a few hundred Myr later in the lower \zstar\ than in the higher \zstar\ models (cf. panels {\it (a)} and {\it (b)} of Fig.~\ref{fig:ionflux}). The photon production rates $Q(\hi)$ and $Q(\hei)$ vary weakly with \zstar\ as compared with its variation with the upper mass cut-off $M_u$. These rates span roughly one order of magnitude from young to older ages in the CSF models explored in this paper. Even though HOLMES provide 100\% of the ionising photons after a few hundred Myr in SSP models, their contribution is negligible in CSF models, in which, the MS and WR stars are always the dominant sources (panel {\it (c)} of Fig.~\ref{fig:ionflux}). Depending on \zstar, the contribution of WR stars to the ionising photon budget can reach up to 25\%.


\subsection{Shock models}
A fraction of the nuclear gas in HII region galaxies may be ionized by both high velocity shocks in the ISM and supernova remnants. It is thus important to understand the contribution of shocks to their emission spectra since the size, luminosity and morphology of the nuclear HII region are sensitive to the ionizing source \citep[see, e.g.,][and references therein]{Moumen19, Rhea23}.

Using {\sc mappings v\footnote{https://mappings.anu.edu.au/code/}} \citep[][]{Sutherland17}, \cite{Alarie19} recalculated the models published by \cite{Allen08}, extending their calculations to lower metallicities by assuming abundances from \cite{Gutkin16}. Their calculations include only fast (200-1000 km s$^{-1}$) shock models, in which the UV and soft X-ray ionizing photons are produced by hot gas cooling behind the shock front. \cite{Alarie19} explored a larger volume of parameter space, resulting in a larger number of models than those computed by \cite{Allen08}.

\subsection{Photoionization models}

We use the {\sc Cloudy} photoinisation code \citep[][c17.03]{Ferland17} to model the emission from gas heated by the massive young (O, B, and Wolf-Rayet) stars and HOLMES (pAGB) present in the \cb SSP and CSF models described above.
We select SEDs ranging in age from 1 Myr to 10  Gyr, totalling 41 SEDs in the case of the SSP and 24 for the CSF models.
In Table\,\ref{tab:pp} we list the physical parameters used in our photoionization models. 

To explore similar physical parameters to the ones used in previous works for SF galaxies \citep[e.g,][and references therein]{Dale06, Kewley01a, Fernandez16, Gutkin16}, we vary the hydrogen density $n_H$ from  $10^{1}$ to $10^{4}$\,\cmc\ in steps of powers of 10.  

We choose to explore regions with a form factor $f_S=1$ \citep{2015Stasinska_aap576}. This value means that the inner radius of the model corresponds to the value one would have obtained for the Strömgren radius, if the region would had been a filled sphere of photoionized gas. This leads to a morphology that is between the plane parallel and the filled sphere cases. The relation between the form factor $f_S$, the ionising photons emitted by the central source $Q(H)$, the hydrogen density $n(H)$ and the ionization parameter\footnote{The ionization parameter $U$ is defined as:
\begin{equation}\label{eq:u}
 U\equiv\frac{Q(H)}{4\pi r^{2}_{o}n(H)c},  
\end{equation}
the dimensionless ratio between the Hydrogen-ionizing photon and the total-Hydrogen densities. In Eq.~(\ref{eq:u})\ $r_{o}$ is the distance between the ionizing source (the stellar population) and the illuminated face of the cloud (in cm), $n(H)$ is the total (ionized, neutral, and molecular) Hydrogen density in cm$^{-3}$, $c$ is the speed of light, $Q(H)$ is the number of Hydrogen ionizing photons emitted by the source in s$^{-1}$. 
} $U$ is taken from \citet{2015Stasinska_aap576}.

The total metal content is chosen to match the stellar metallicity. 
The abundance of the solar nebula is adopted from \citet[][]{Bressan12}, as in \citet[][]{Gutkin16}\footnote{The initial metallicity of the solar nebula corresponds to $Z = 0.017$, and the current surface abundance of the sun to $Z$\,=\,0.015 \citep[][and references therein]{Gutkin16}}.
We allow C/O and N/O to vary (see Table\,\ref{tab:pp}). Once the values for the total abundance ratios of O/H, N/O and C/O are defined, we obtain the gas abundance by applying a depletion to every element, using the recipes given by \citet{Gutkin16}, with a dust-to-metal mass ratio parameter $\xi_d = 0.36$. 

We include dust grains using the "ism" default dust composition in {\sc Cloudy}. The amount of dust in the {\sc Cloudy} models is determined by scaling the solar value by $\xi_d / 0.36 \times Z  / 0.015$, where Z is the total metallicity.

Finally, we take several cuts in the spatial integration of the column density in the models. All {\sc Cloudy} photoionization models are obtained using a stopping criteria of H$^+$/H = 0.02. No electron temperature stopping criteria is used, to avoid metal-rich models to artificially stop before reaching the recombination front. This leads to radiation-bounded (RB) models at all metallicities. From each individual model, we generate 4 other models by considering the values of the output variables (line intensities, mean values of physical parameters, ionic fractions, etc.) obtained after cutting the initial model at radii corresponding to $\hb =$ 80, 60, 40, and 20\% of the total RB-case \hb\ value. 

In total, we generated 2,358,330 models, resulting from exploring two SF laws, two IMFs, two upper mass cut-offs, seven stellar and nebular metallicities, 24 ages for the SSP, 41 ages for the CSF case, four electron densities, six ionization parameters, three C/O ratios, three $\Delta$N/O departure from canonical N/O value, and six values for the cut in radius (see above). 

We note that there is a large number of RB models in the normal-SF zone. These models can be associated with optically thick clouds, in which the emission of \neiiw\ is efficient \citep[][]{Melendez14}.
For each model we compute the principal optical and IR emission lines typically used to classify galaxies. 

The models presented in this section are available\footnote{\url{https://sites.google.com/site/mexicanmillionmodels/the-different-projects/cb_19?pli=1}} in the Mexican Million Model database \citep[3MdB][]{2015Morisset_rmxa51}
\footnote{\url{https://sites.google.com/site/mexicanmillionmodels/}}
under the reference code CB\_19. The matter-bounded (MB) models can be identified by their HbFrac parameter value. 

\section{IR emission line ratios}

Since dust extinction decreases with wavelength, IR emission lines are especially suited to explore heavily obscured galaxy nuclear regions, where copious star formation and accretion onto supermassive black holes takes place.
The emission from the gas excited by hot stars provides information about both the stellar population and the surrounding gas. 
In this section we use the results of Sec.~\ref{sec:models} to explore the IR emission line ratios used frequently in the literature to classify low-luminosity galaxies and/or deeply obscured systems that go undetected in the visible range. 

\begin{figure*}
  \centering
  \includegraphics[width=0.820\textwidth]{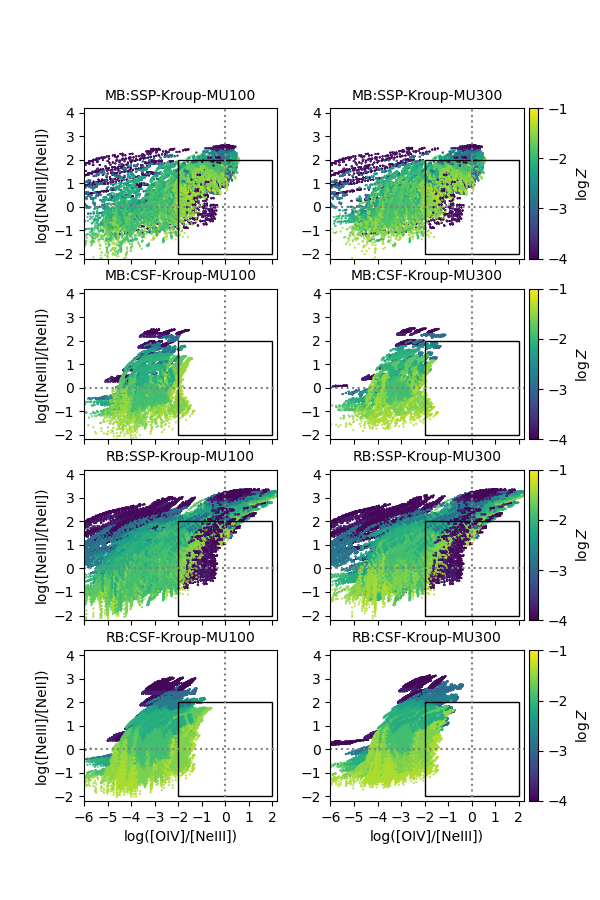}
  \caption{Synthetic $\log\,(\neiiiw/\neiiw)$ vs. $\log\,(\oivw/\neiii)$ MIR diagnostic diagram from the SSP and CSF models assuming a Kroupa IMF with either $m_u$\,=\,100 or 300 $\Msol$. Models reaching the MB condition are shown in the {\it 4-top panels}. The {\it 4-bottom panels} show the RB models. Models are colour codded according to $\log\,Z$ as indicated in the colour-bar. We remark that the range of values of both line ratios plotted in these frames is considerably larger than the range of interest shown in Fig.\,\ref{fig:DD_def}, indicated by the inner square in each panel.}
 \label{fig:kroupa_mir}
\end{figure*}

\begin{figure*}
  \centering
  \includegraphics[width=0.825\textwidth]{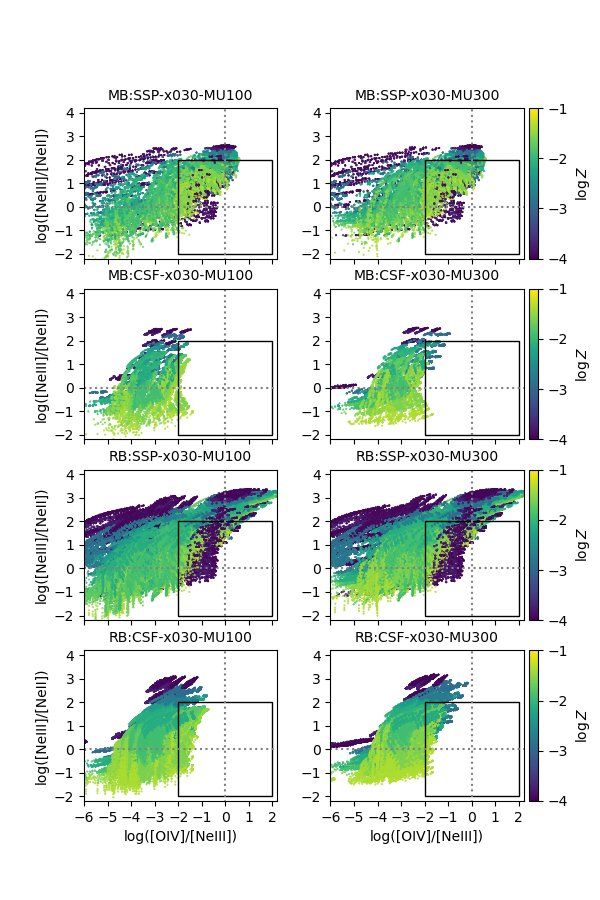}
  \caption{Same as Fig.~\ref{fig:kroupa_mir} but for the x030 IMF.}
  \label{fig:x030_mir}
\end{figure*}

In Fig.~\ref{fig:kroupa_mir} we plot the resulting MIR diagnostic diagram for the SSP and CSF Kroupa IMF models for both $m_u$\,=\,100 and 300 $\Msol$. Models reaching the MB condition are shown in the {\it top panels}. The {\it bottom panels} show the RB models. MB and RB models simulate partially and totally ionized clouds, respectively.
Models are colour-coded according to their metallicity. 
Fig.~\ref{fig:x030_mir} shows a similar plot but for the x030 IMF models.
There are no significant differences between equivalent panels in
Figs.~\ref{fig:kroupa_mir} and ~\ref{fig:x030_mir}, implying that neither MB nor RB model predictions are sensitive to the IMF slope or the upper mass cut-off. This behaviour is expected since the parameters that affect most the SEDs are the age and the metallicity (cf. Fig.~\ref{fig:ionphot}). Since the IMF slope and the upper mass cut-off do not affect significantly the ionizing SED, the properties of the ionized gas do not depend on these parameters.

The same behaviour is observed in the optical spectral range, see Figs.~\ref{fig:kroupa_opt} and \ref{fig:x030_opt} in Appendix~\ref{AppxA}. 
In the same token, \citet[][]{Gutkin16} successfully modelled observations of star forming galaxies in several optical diagnostic diagrams, using a similar grid of stellar population models and gas cloud parameters. They also find marginal differences between the line emission ratios expected from models with different IMFs and upper mass limits (see their figure 6), as we do for the IR lines.

\begin{figure*}
  \centering
  \includegraphics[width=1.1\textwidth]{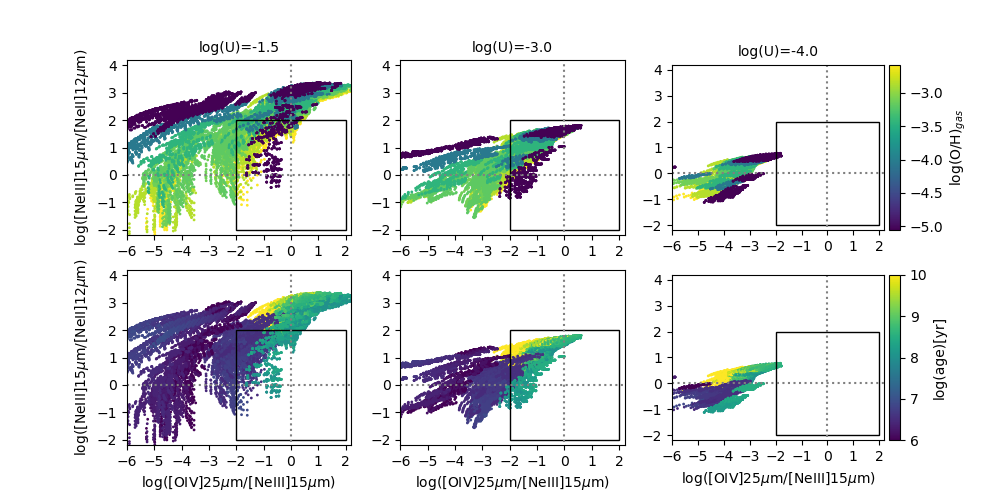}
  \caption{Synthetic $\log\,(\neiiiw/\neiiw)$ vs. $\log\,(\oivw/\neiii)$ MIR diagnostic diagram for the RB-SSP-Kroup-MU100 models. {\it Left panels:} \logu\,=\,-1.5. {\it Middle panels:} \logu\,=\,-3. {\it Right panels:} \logu\,=\,-4. Points are coded in size according to $\log{n_{e}}$ and colour coded according to log(O/H) (top panels) and age (bottom panels), as indicated by the colour bars.}
  \label{fig:para}
\end{figure*}

In Fig.~\ref{fig:para} we explore the dependence of our models on $U$. We plot the MIR diagnostic diagram for three values of $U$, spanning the full range used in this investigation. A fraction of the \logu\,=\,-1.5  (left panel) models has $\log\,(\oiv/\neiii)$\,$>$\,0, while most \logu\,=\,-3 (middle panel) and all \logu\,=\,-4 (right panel) models have $\log\,(\oiv/\neiii)$\,$<$\,0.
It is important to note that the range of values of the \oiv/\neiii\ ratio allowed by the models is considerable larger than observed, as highlighted by the inner square in each panel of Figs.~\ref{fig:kroupa_mir}\,-\,\ref{fig:para}. 
The low $\log\,(\oiv/\neiii)$ values, reaching -6, imply that there is no \oiv\ emission in the cloud.

For $\logu\,<\,-3$ there is virtually no O$^{+3}$ emission because the radius-averaged ionization fraction decreases, making the gas inefficient in producing \oiv\,\citep[][figure 13]{Melendez14}.
Despite these extremely low values of \oiv/\neiii\, the values of 
\neiii/\neii\ are still relevant in a Ne dominated gas, and can still be used as a single-ratio diagnostic in the absence of \oiv\ emission in SF galaxies.

The SSP models allow for a wider range of values of the $\oiv/\neiii$ and $\neiii/\neii$ line ratios and offer a larger baseline to understand the complex nature of SF regions in galaxies, in which low and highly ionized atomic species coexist. From here on and for clarity, we will discuss only the results for the RB, \mup\,=\,100\,\Msol\ Kroupa IMF SSP models, labelled (RB-SSP-Kroup-MU100) in the plots. Similar conclusions will be reached using the other sets of SSPs, including the MB models. \citet[][]{Gutkin16} assumed (for simplicity) that galaxies are ionization-bounded and therefore used only RB models (labelled ionization-bounded models in their paper).

\begin{figure*}
  \centering
  \includegraphics[width=0.90\textwidth]{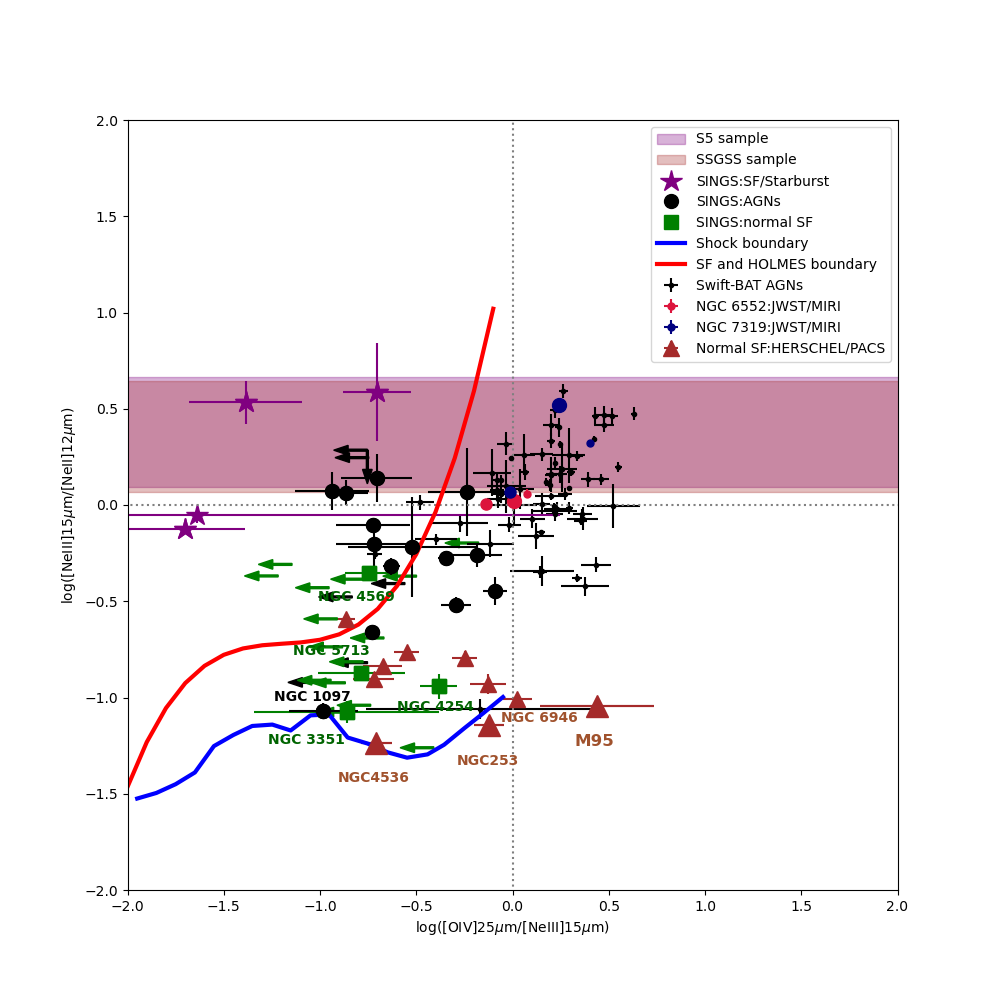}
  \caption{$\neiiiw/\neiiw$ vs. $\oivw/\neiiiw$ MIR diagnostic diagram. The {\it red} line represents the boundary of the SF-zone defined in Eq.~(\ref{eq:single_zone_mir}). The {\it blue} line represents the lower-limit for the excitation from shock models in the normal-SF zone.
  The meaning of the various symbols is explained in the legend. The black and green arrows indicate upper limits. {In the case of NGC~6552 the size of the {\it crimson} points corresponds to the nuclear (small), circumnuclear (medium), and central (large) emissions. In the case of NGC7319 the size of the {\it navy} points correspond to the nuclear (small), N2 (medium), S2 (large). Larger triangles are used to highlight individual objects.}}
  \label{fig:mirb}
\end{figure*}

\begin{figure*}
  \centering
  \includegraphics[width=0.60\textwidth]{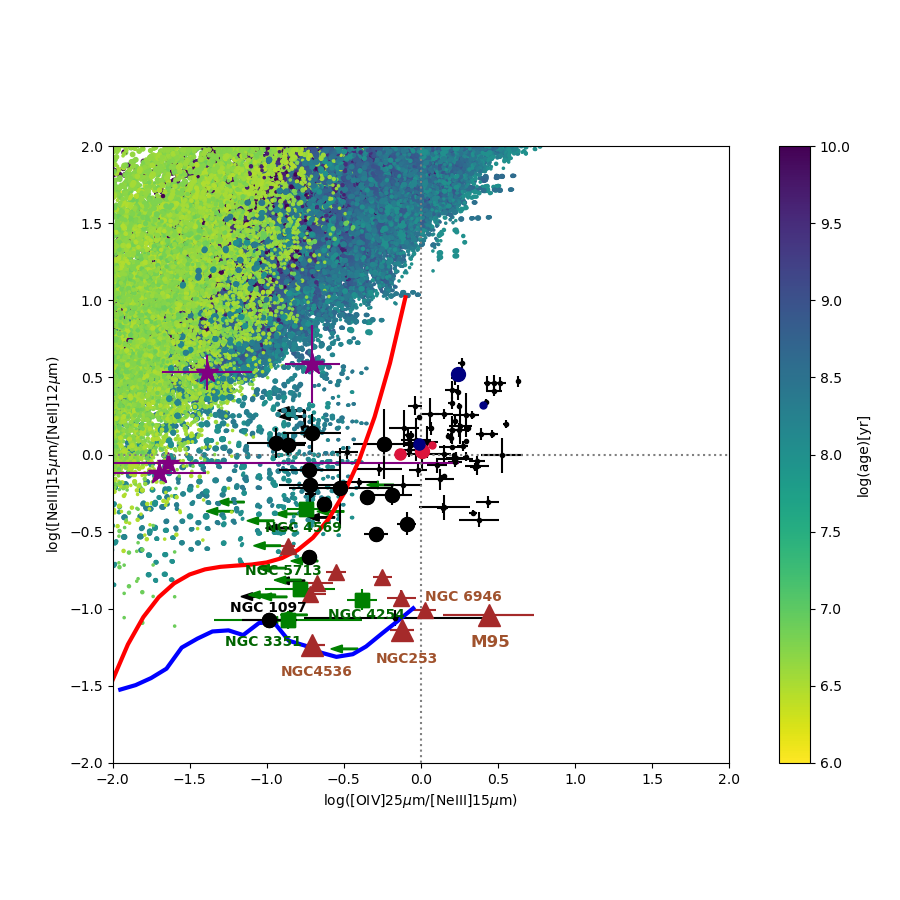}
  \includegraphics[width=0.60\textwidth]{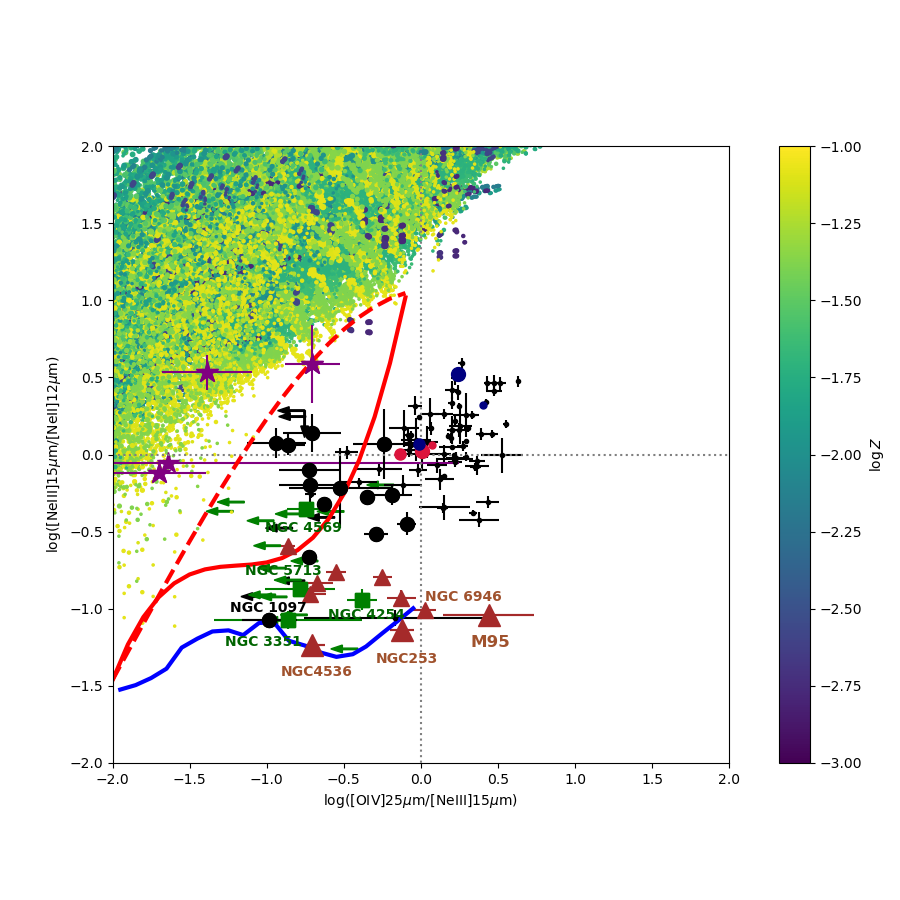}
  \caption{$\neiiiw/\neiiw$ vs. $\oivw/\neiiiw$ MIR diagnostic diagram. {\it Upper panel:} the {\it red solid} line represents the boundary of the SF-zone defined in Eq.~(\ref{eq:single_zone_mir}). The models are colour coded by age according to the colour bar. {\it Bottom panel:} the {\it red dashed} line represents the boundary of the SF-zone defined in Eq.~(\ref{eq:single_zone_mir_disscusion}). The models are colour coded by metallicity according to the colour bar. In each panel the {\it blue} line represents the lower-limit for the excitation from shock models in the normal-SF zone. The meaning of the various symbols is explained in the legend in Fig.~\ref{fig:mirb}. The black and green arrows indicate upper limits. Larger triangles are used to highlight individual objects.}
  \label{fig:mir}
\end{figure*}

To explore how model predictions overlap with the observations, in Figs.\,\ref{fig:mirb} and \ref{fig:mir} we compare our results with the sample discussed in Sec.~\ref{sec:sample}\,(Table\,\ref{tab:observations}). The cubic relation
\begin{equation}
\begin{split}
	\log\,(\neiii/\neii)\ &=\ 5.6\times \log\,(\oiv/\neiii) \\
	& +\ 4.6\times \log\,(\oiv/\neiii)^{2} \\
	& +\ 1.3\times \log\,(\oiv/\neiii)^{3} + 1.5,\\
 {\rm for} &\ \ \ -2 \lesssim \log\,(\oiv/\neiii) \lesssim -0.05, \\
\end{split}	
\label{eq:single_zone_mir}
\end{equation}
({\it red solid} line in top panel of Fig.\,\ref{fig:mirb} and \ref{fig:mir}) represents the lower boundary of a region that encloses models of age from 100 to 1000\,Myr, falling in quadrants II and III of Fig.\,\ref{fig:DD_def}. The fit in Eq.~(\ref{eq:single_zone_mir}) excludes the extreme models with $\log\,(\oiv/\neiii)$\,$\gtrsim$\,0, with parameters \logu\,=\,-2 and 10\,$\leq$\,$n_{H}$\,$\lesssim10^{3}$\,\cmc. For comparison, models with $\log\,(\oiv/\neiii)$\,$\sim$\,0 have \logu\,=\,-3 for all values of $n_H$. 

The top panels of Fig.~\ref{fig:para}, show that only the lower metallicity models fill quadrants II and III in the range of interest shown in Fig.\,\ref{fig:DD_def}, whereas models of higher metallicity fall in quadrant III of the diagram but outside this range.
The latter being dominated by stellar populations younger than 1 Gyr, with older populations having the lower metallicities (bottom panels of Fig.~\ref{fig:para}). 
By excluding the lowest metallicity models (log\,$Z<-3$), we derive a second star forming boundary,  described by the following cubic relation
\begin{equation}
\begin{split}
	\log\,(\neiii/\neii)\ &=\ 0.1\times \log\,(\oiv/\neiii) \\
	& -\ 0.9\times \log\,(\oiv/\neiii)^{2} \\
	& -\ 0.2\times \log\,(\oiv/\neiii)^{3} + 1.1,\\
 {\rm for} &\ \ \ -2 \lesssim \log\,(\oiv/\neiii) \lesssim -0.05, \\
\end{split}	
\label{eq:single_zone_mir_disscusion}
\end{equation}
and shown as a {\it red dashed} line in the bottom panel of Fig,\,\ref{fig:mir}. Interestingly, SINGS AGNs are below this boundary. 

Shocks are an important source of ionization in objects in quadrant III of the MIR diagnostic diagram. Shocks can be produced by either cloud-cloud collisions, the expansion of HII regions into the surrounding interstellar medium, outflows from young stellar objects, supernova blast waves, outflows from AGNs and starbursts, and/or galaxy collisions.

\begin{figure*}
\centering
\includegraphics[width=0.90\textwidth]{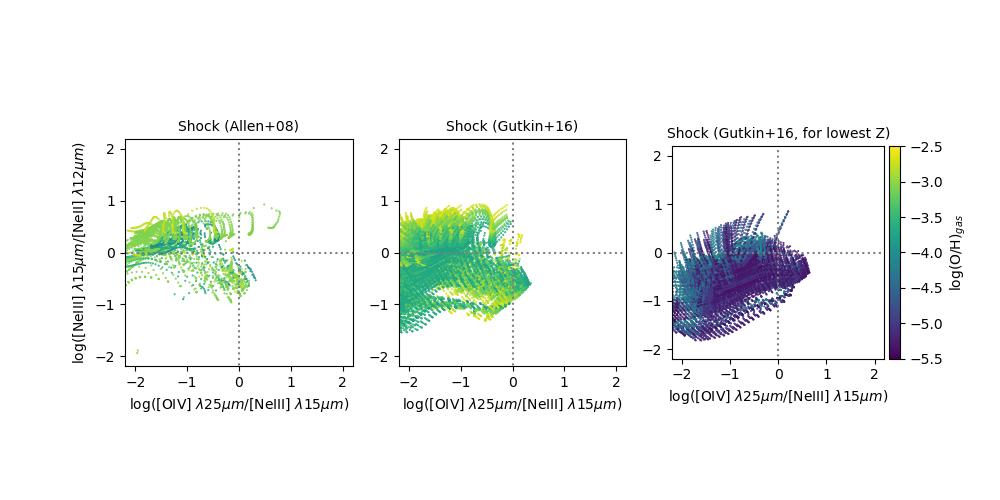}
\caption{Position of shock models in the $\neiiiw/\neiiw$ vs. $\oivw/\neiiiw$ diagnostic diagram. {\it Left panel:} shock models computed by 
\citep{Allen08} using {\sc mappings v} 
\citep{Alarie19}. {\it Middle panel:} shock 
models computed by \citep{Allen08} using {\sc mapping v} and assuming the abundances from 
\citep{Gutkin16}. {\it Right panel:} the same as 
in the {\it middle panel} but only for the lower 
metallicities. Note that we are plotting the 
Shock+Precursor models from \citep{Alarie19}, 
available in the 3MdB database 
\citep{2015Morisset_rmxa51}. Points are colour 
coded according to log\,($O/H$), as indicated in 
the colour bar.
} 
\label{fig:shocks}
\end{figure*}

We observe a considerable increase in the number of shocks models calculated by \citet{Alarie19} lying in quadrant III. Excluding models with the lowest metallicity, 
we derive the lower-limit for excitation produced by shocks, indicated by the {\it blue solid line} in Figs.\,\ref{fig:mirb}\,and\, \ref{fig:mir}.
In Fig.~\ref{fig:shocks} we plot results for shock+precursor models, which take into account the gas entering the shock front.

We highlight that it is possible that a fraction of the observed MIR emission lines in \hii\ galaxies is due to collisional excitation by shocks \citep[e.g.,][]{Groves04}. It is likely that in some HII galaxies shocks driven by a weak AGN-outflow are an important source of excitation \citep[e.g., ][]{Cazzoli22}. For the three galaxies (NGC~253, NGC~6946, and M~95) below the lower boundaries for excitation from our photoionization ({\it red-solid} and {\it red-dashed} lines) and shock ({\it blue-solid} line) models in Figs.\,\ref{fig:mirb} and \ref{fig:mir}, it is likely that another ionising source, like a weak AGN, needs to be considered.

\cite{Feltre23} find that in Seyfert galaxies in which the contribution from the AGN is below $50\%$, MIR line ratios are better modelled when shocks are added as an additional source of ionization .
Recent studies have shown that the star formation activity in galaxies is highly concentrated towards their centre \citep{Martinez_Paredes19}.
Additionally, there is growing evidence indicating that the AGN dusty torus is a structure that includes a polar wind, that originates in the inner part of the torus and that is co-spatial with the outflow
\citep[e.g.,][and references therein]{Martinez_Paredes20,Garcia_Bernete22}. Thus, we cannot neglect the possibility that shocks produced by stellar feedback are also an important source of ionization, especially in objects that lie close to the star forming boundary.

In general, the star forming activity of  galaxies in quadrant III
must occur under very special conditions, in which shocks and AGNs influence their evolution. A deeper study to disentangle their different contributions on scales of a few parsec can be carried out with {\it JWST}/MIRI, which offers the needed combination of high spatial resolution and sensitivity. 


\begin{figure*}
  \centering
  \includegraphics[width=0.60\textwidth]{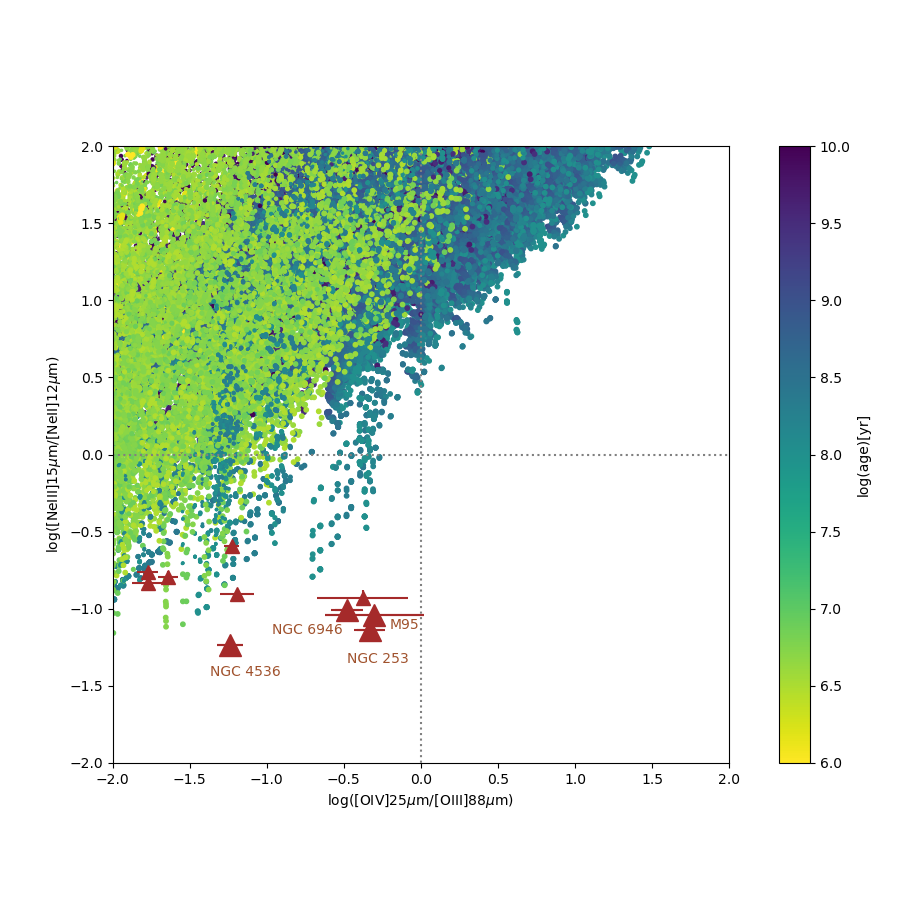}
  \caption{$\neiiiw/\neiiw$ vs. $\oivw/\oiiiw$ MIR-FIR diagnostic diagram proposed by \citet[][]{Fernandez16}. The {\it triangles} show the \citet{Fernandez16}
  {\it Herschel}\,/\,PACS observations listed in Table~\ref{tab:observations}.
  Models are colour coded according to log age as indicated in the colour bar. Larger triangles are used to indicate individual objects.}
  \label{fig:fir}
\end{figure*}
\begin{figure*}
\centering
  \includegraphics[width=1\textwidth]{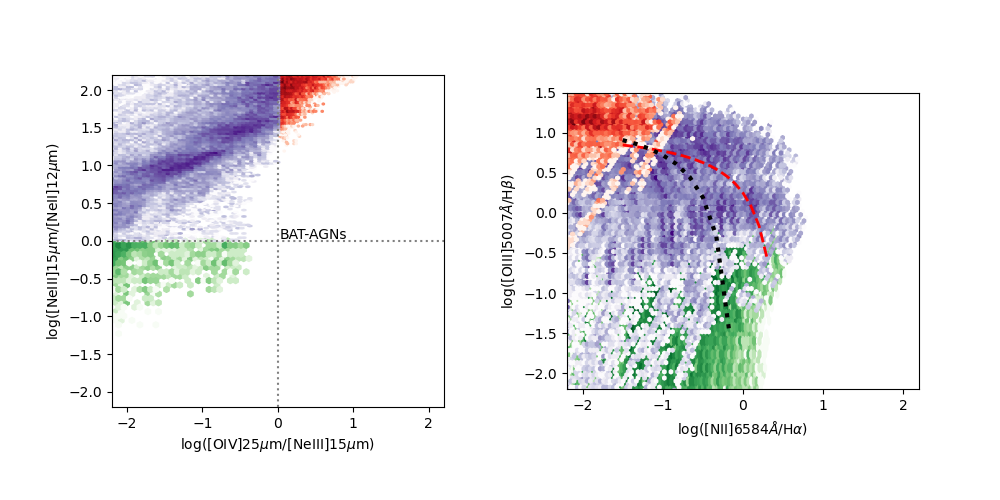}
  \caption{$\neiiiw/\neiiw$ vs. $\oivw/\neiiiw$ MIR diagnostic diagram ({\it left}) and
  BPT diagram ({\it right}). Models that in the MIR diagram lie in the quadrant III, quadrant II, and quadrant I are plotted in {\it green, blue} and {\it red}, respectively, in both diagrams. 
  The HDL defined in Eq.~(\ref{demarc_line}) is plotted as a {\it red-dashed} line in the BPT diagram. The {\it black-dotted} line is the minimal AGN contribution from \citet[][]{Kauffmann03}.}
  \label{fig:regions}
\end{figure*}

\begin{figure*}
  \centering
  \includegraphics[width=0.90, width=\textwidth]{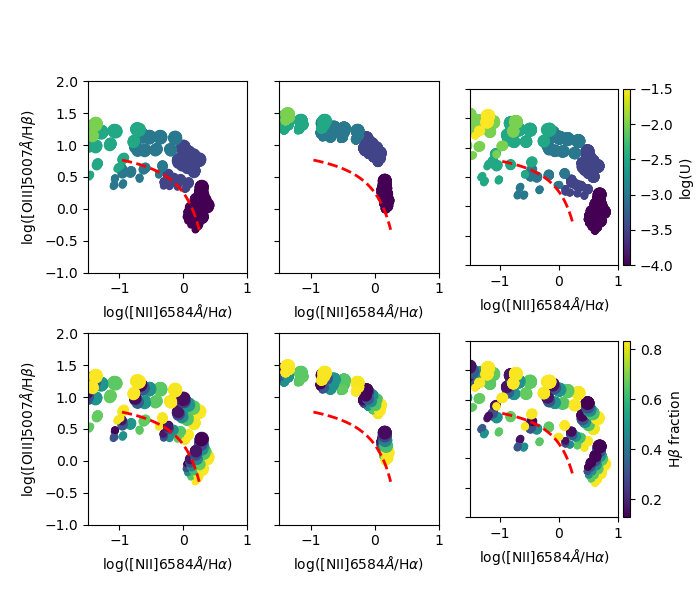}
  \caption{Position of the RB-SSP-Kroup-MU100 models with extreme values of parameters in the BPT diagram. Models are shown at age\,=\,1 Gyr for
  $\Delta N/O$\,=\,0, $C/O$\,=\,-0.36 ({\it left}),
  $\Delta N/O$\,=\,-0.25, $C/O$\,=\,-1 ({\it middle}) and
  $\Delta N/O$\,=\,+0.25, $C/O$\,=\,-1 ({\it right}).
  In the upper panels the points are colour coded according to \logu, as indicated in the color bar. In the bottom panels the points are colour coded according to $H_{\beta}$ fraction. In each panel the size of the points increases with $n_{e}$ and the HDL, defined by Eq.~(\ref{demarc_line}), is plotted as a {\it red-dashed} line.}
  \label{fig:opt_grid}
\end{figure*}

\begin{figure*}
  \centering
  \includegraphics[width=0.90\textwidth]{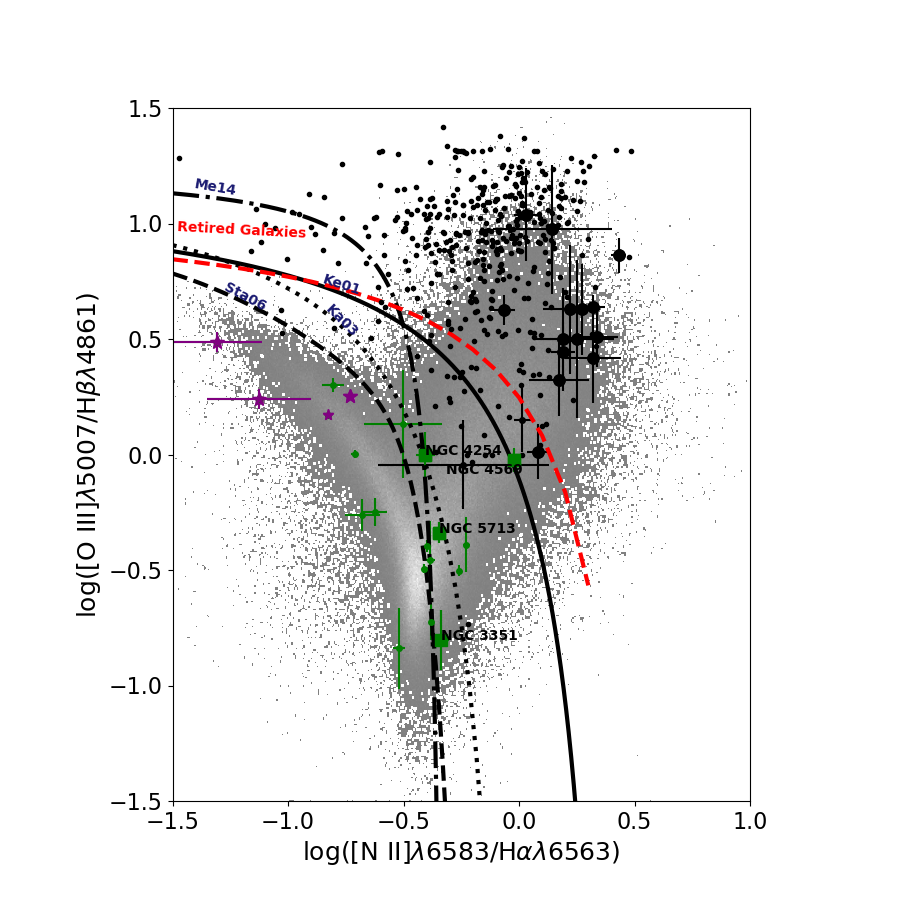}
  \caption{Position of various demarcation lines in the BPT diagram.
  The {\it red-dashed} line represents the HDL defined in Eq.~(\ref{demarc_line}).
  The {\it black-dot-dashed} line represents the minimum AGN contribution derived by \citet[][Me14]{Melendez14}.
  The {\it black-dashed} line is the theoretical limit for the pure SF galaxies from \citet[][Sta06]{Stasinska06}.
  The {\it black-solid} line is the theoretical maximum contribution from SB calculated by \citet[][Ke01]{Kewley01}.
  The {\it black-dotted} line is the minimal AGN contribution from \citet[][Ka03]{Kauffmann03}.
  The meaning of the {\it purple, green} and {\it large-black} symbols is the same as in Fig.~\ref{fig:mirb}. The {\it small-black} points are from the BAT-AGN sample \citep[][]{Koss17}.
  The {\it grey-dots} are galaxies from the SDSS DR8 selected with signal-to-noise 
  ratio\,>\,3\,$\sigma$ \citep[see][]{Brinchmann04,Tremonti04}.}
\label{fig:opt_obs}
\end{figure*}


\subsection{Discussion}

Most of our models in quadrant III with the higher ratios of $\oivw/\neiiiw$ have low metallicity. However, HII galaxies in our sample have nuclear or central metallicities not far from solar abundance, with (O/H) around -3.5 to -3 \citep[][]{Moustakas10, Beck22, Kaplan16}. These metallicities have been derived from optical spectra using different methods, for example, strong-line and radial gradients \citep[][]{Moustakas10}. \citet{Beck22} used the MIR spectrum and the equations from \citet{Nicholls17}.

Due to the presence of different sites of star formation within the central region of these galaxies, it is likely that some of them have a complex distribution of metallicity \citep[e.g.,][]{Calzetti21, Boselli06, Barth00}. In particular, NGC~3351 has both a circumnuclear starburst ring and a nuclear red stellar population within the first kiloparsec \citep[][]{Calzetti21}. For NGC~253, metallicities ranging from $\sim$\,0.5\,$Z_{\odot}$ \citep[][from the X-ray spectrum]{Ptak97} to $\lesssim$\,1.5\,$Z_{\odot}$ \citep[][from the optical spectrum]{Webster83}, have been reported. 

In NGC~4569, a prototypical anaemic galaxy located close to the center of the Virgo Cluster, the observed nuclear optical spectrum is better modelled under a ram pressure stripping scenario. NGC\,253 and NGC\,4536 show some evidence of a weak AGN \citep[e.g.,][]{Hughes05,Mohan02}. In this case, \oivw\ turns out to be a good indicator of AGN activity, since the emission of this line increases in presence of an AGN radiation field \citep[e.g.,][]{Melendez08a}.

Among the \hii\,galaxies in this sample, NGC~3351 and M95 have been identified as hot-spot galaxies, in which the central source, a LINER or a Seyfert nucleus, is surrounded by an HII region within the first kiloparsec \citep[]{Sersic73, Kennicutt89}. Curiously, the active galaxy NGC~1097, also identified as a hot-spot galaxy, has similar MIR emission line ratios to NGC~3351, while M95 presents the highest $\log\,(\oivw/\neiiiw)$ line ratio. On the other hand, NGC 4569 has been classified as a transition object \citep[LINER/HII,][]{Keel96, Ho97b}, in which 
the UV and optical properties are attributed to the presence of a young in combination with an old stellar population or AGN. However, based on the position of this galaxy in the MIR and optical BPT diagrams, the dominant ionising source thought to be HOLMES+AGN 
might be important in sources like NGC\,5713, where many nuclear HII regions have been detected, and NGC\,4254, where the spiral structure is very complex and chaotic \citep[]{Eskridge02,Martini03,Brookes06}.

In Figs.~\ref{fig:mirb} and \ref{fig:mir} we include the nuclear and extended emission from the nearby AGNs NGC~6552 (crimson dots) and NGC~7319 (navy dots), recently observed with the {\it JWST}/MIRI instrument. Both objects have been identified as obscured-AGNs, with $\log{N_H}$\,=\,24.05 and 23.8 [cm$^{-2}$], respectively.
NGC~6552 has been optically classified as Seyfert 2 \citep[][and references therein]{Alvarez_Marquez22}.
NGC~7319 is a type 2 AGN which lacks a nuclear starburst \citep[][and references therein]{Pereira-Santaella22}. Additionally, it has two asymmetric radio lobes and a compact core, likely associated with the presence of a radio jet. It shows an outflow that is co-spatial with the radio lobes, suggesting that the outflow is driven by the radio jet \citep[e.g,][]{Aoki99}.

For NGC~6552, we plot the emission of the nuclear (within $\sim$\,0.58\,kpc, small crimson dot), circumnuclear (between $\sim$\,0.55 and $\sim$\,0.88\,kpc, medium crimson dot), and central (large crimson dot) regions. The nuclear and central emissions lie in the AGN zone while the circumnuclear emission shifts towards the star forming zone. \cite{Alvarez_Marquez22} point out that the central spectrum includes the emission of the nuclear and circumnuclear regions, but it is not the sum of both of them. In the case of NGC~7319, we plot the emission from the nuclear AGN (small navy dot), from the radio hotspot N2 (medium navy dot), and from the radio hotspot S2 (large navy dot). The nuclear and S2 (located at 1.5 kpc from the unresolved radio source) emissions clearly lie in the AGN zone. However, N2 (located at 430 pc from the unresolved radio source) shifts towards the transition zone. \cite{Pereira-Santaella22} argue that this asymmetry can be explained by the interaction of the northern radio jet (where the N2 hotspot is located) with the ISM. These results illustrate the power of the MIR diagram to classify the different excitation mechanisms present in the nuclear and circumnuclear regions of galaxies.

In Fig.~\ref{fig:fir} we plot the \neiii/\neii\,vs.\,\oiv/\oiii\ MIR-FIR diagram proposed by \citet[][]{Fernandez16} as a diagnostic diagram for SB and AGN activity. Our photoionization models reproduce about half of their sample, most likely as a result of the larger number of stellar parameters explored by us (e.g., metallicity, IMF, and age). We note that in this diagram, NGC\,4536, NGC\,253 and M95 lie outside of the zone covered by our photoionization models. Comparing the physical conditions in different \hii\,regions, \citet[]{Kennicutt89} concluded that in \hii\,galaxies with peculiar properties (e.g., broader and extended \oiiiwv\ emission) like NGC\,3351, a fraction of the observed line emission could be attributed to shocks or to an unresolved LINER or Seyfert nucleus surrounding an SF region. Incidentally, \citet{Ho97} found that in early-type spiral galaxies, like many \hii\,galaxies, mass loss from evolved stars in the central $\sim$\,5\,pc can provide most of the fuel required by the AGN. Recent studies have found that nuclear SB can survive the strong radiation field of the AGN down to scales of a few parsec in the case of Seyfert galaxies \citep{Esquej14} and of a few hundred parsecs in the case of quasars \citep{Martinez_Paredes19}.

\section{BPT diagram}
The classical BPT diagnostic diagram \citep{BPT81} is an important tool to discriminate between SF and AGN-dominated emission regions. However, since optical emission lines are affected by extinction, its classifying power decreases for low-luminosity and optically-obscured systems. In general, the optical spectra of SF galaxies (like WR and BCD galaxies) can be described by a population of Wolf-Rayet and massive hot OB stars, since they are the main source of UV radiation. However, in other galaxies, an additional ionising source, either UV radiation produced by a population of HOLMES or shocks, seems necessary to explain the observed emission line ratios \citep{Feltre23}. In general, authors tend to use shock models to explain the excess of emission that cannot be reproduced by massive hot stars \citep[e.g.,][and references therein]{Flores11}. However, studying the optical and MIR colours of a sample of LINER-like galaxies, \cite{Flores11} found that in these galaxies HOLMES are an important source of ionization.
 
In Fig.~\ref{fig:regions}, we plot next to each other the MIR and BPT diagnostic diagrams for the RB-SSP-Kroup-MU100 photoionization models, colour-coded according to their position in the MIR diagram. Models  ({\it green} dots) in quadrant III of the MIR diagram reach lower values of $\oiii/\hb$ than those in quadrants I and II ({\it blue} and {\it red} dots). However, most of these models have $\oiv/\neiii$ lower than -2. This may explain why there are so few {\it pure} SF galaxies in this zone with detected $\oivw$ emission. The detection of $\oiv$ in galaxies that lie in this zone might indicate the presence of an additional source of excitation, like shocks driven by a weak AGN. On the other hand, a number of models in the HOLMES regime reach $\oiii/\hb$ ratios larger than in SB models. Using optical and MIR data for a LINER sample, \cite{Herpich16} found that a non-negligible population of LINERs are retired galaxies, in which the star formation activity has stopped, giving pass to ionization powered by HOLMES.

Following \citet[][]{Kewley01} definition of a {\it maximum SB line}, we use the subset of our RB-SSP-Kroup-MU100 models with typical parameters  $\log\,(OH)_{gas}$\,=\,-2.86, $\Delta N/O$\,=\,0, $C/O$\,=\,-0.36, $n_{e}$\,=\, 10$^{2}$\,\cmc, age\,=\,1\,Gyr
\citep[e.g.,][]{Flores11, Sokolowski91}, to define a new optical 
HOLMES {\it demarcation line} (HDL), described by the following fitted relation
\begin{equation}
    \log\,(\oiii/\hb) < \frac{0.51}{\log\,(\nii/\ha)-0.61}+1.09,
    \label{demarc_line}
\end{equation}
representing the  
{\it maximum} contribution by HOLMES to the excitation in retired galaxies,
avoiding the AGN-dominated region in the MIR diagnostic diagram.
The HDL is plotted as a dashed line in the BPT diagram shown on the RHS panel of Fig.~\ref{fig:regions}. 

In Fig.~\ref{fig:opt_grid} we plot models corresponding to extreme values in our grid of parameters, together with the HDL. These extreme models are among the {\it red} points in Fig.~\ref{fig:regions}.
In Fig.~\ref{fig:opt_grid} models are shown at age\,=\,1\,Gyr for $n_{e}$ in the range $10-10^{4}$\,\cmc.
Models in the LHS panels, corresponding to
($\log\,(OH)_{gas}$\,=\, -2.86, $\Delta N/O$\,=\,0, $C/O$\,=\,-0.36),
with $n_{e}$\,>\,$10^{2}$\,\cmc exceed the HDL limit and mimic the radiation 
field in AGN-dominated sources \citep{Stasinska15}. 
Similarly, models in the middle
($\log\,(OH)_{gas}$\,=\, -2.58, $\Delta N/O$\,=\,-0.25, $C/O$\,=\,-1),
and RHS 
($\log\,(OH)_{gas}$\,=\, -2.58, $\Delta N/O$\,=\,0.25, $C/O$\,=\,-1) panels exceed the HDL limit for all values of $n_{e}$. 
Most likely, these models overestimate the excitation because they are calculated assuming extreme values of $(OH)_{gas}$ and $\Delta N/O$ \citep[see Fig. 1 in][]{Gutkin16}. Therefore, the physical conditions of the models in the left panels of  Fig.~\ref{fig:opt_grid} with $n_{H}=10^{2}$ \cmc\ represent a more realistic scenario to model photoionization by HOLMES in retired galaxies.

In Fig.~\ref{fig:opt_obs} we plot the HDL, Eq.~(\ref{demarc_line}), in the BPT diagram, together with AGN ({\it black}), SB ({\it purple}) and \hii\ galaxy ({\it green}) observations. It is likely that AGNs that lie below the HDL are contaminated by emission due to HOLMES.
Interestingly, whereas in the BPT diagram HII galaxies lie within the minimal AGN boundary \citep[lines marked Ka03 and Me14 in Fig.~\ref{fig:opt_obs};][]{Kauffmann03,Melendez14} 
and the maximum stellar boundaries \citep[line marked Ke01;][]{Kewley01} and the HDL, 
in the mid-IR diagram most of them lie out of the star forming boundary, indicating that shocks, attributed to stellar feedback or/and AGN activity, are an important source of excitation. This result shows the robustness of using both the optical and IR diagnostic diagrams to classify deeply dust-enshrouded and obscured systems. A better comprehension of the ionising sources translates into a deeper understanding of the evolutionary path of these galaxies.

We remark that our SB models with $n_{H}$\,=\,$10^{2}$ \cmc\ are in agreement with the theoretical maximum SB line derived by \citet[][]{Kewley01}
using the {\sc PEGASE v2.0} \citep[][]{Fioc97} and {\sc Starburst99} \citep[][]{Leitherer99} stellar population models and the {\sc mappings iii} \citep[][]{Sutherland93} photoionization code, for $n_{H}$\,=\,$10^{2}$\,\cmc\ and a maximum stellar age of 100\,Myr. More complete studies with optical and IR samples of AGN and retired galaxies are needed to fully understand the different nature of their ionizing radiation.

\section{Summary and conclusions}\label{sec:conclusions}

In this paper we investigate the excitation mechanisms and the physical properties of the ionized gas of star forming galaxies appearing in the  $\neiii/\neii$\,vs.\,$\oiv/\neiii$ MIR diagnostic diagram proposed by \citet[][]{Weaver10}. This MIR diagram is divided into four zones according to the value of the emission line ratios. Using the \cb\ stellar population models described in \citet{plat2019} and 
\citet{Sanchez22}, together with the {\sc Cloudy} photoionization code  \citep[][]{Ferland17}, we calculate the intensity of the following optical, MIR and FIR emission lines: \oiiiwv, \hb, \niiw, \ha, \oivw, \neiiiw, \neiiw\ and \oiiiw, expected to be present in the spectra of SF galaxies. 

We explore \cb\ models forming stars either in an instantaneous burst (SSP) of mass $1\times10^{6}$\,\Msol\ or at a constant rate of 1\,\Msol\,$yr^{-1}$ (CSF). We compute these models for the Kroupa \citep{Kroupa01} and a top-heavy ($x$\,=\,0.30) IMF with two upper mass limits, $M_{u}$\,=\,100 and 300\,\Msol. We assume seven stellar metallicities in the range $0.0002<\zstar<0.06$. To model both star forming and retired galaxies, we allow the age of the stellar population to vary from 1 Myr to 10 Gyr.
For the photoionized clouds we vary the ionization parameter $\log(U)$, the hydrogen density $n_H$, and $\log\,(O/H)_{gas}$, $\Delta N/O$ and $C/O$ as indicated in Table~\ref{tab:pp}. The gas metal content is assumed to be the same as that of the ionising stars.
We find that, in general, the MIR ($\neiii/\neii$\,vs.\,$\oiv/\neiii$) line ratios are insensitive to variations in the IMF slope and upper mass cut-off. Althought most models lie out of the range of interest (quadrant III in Fig.\,\ref{fig:DD_def}), a minor fraction falls in this quadrant. SSP models span a wider range of $\neiii/\neii$\,vs.\,$\oiv/\neiii$ line ratios. Therefore, for simplicity, we base our analysis on radiation bounded (RB) models with \mup\,=\,100\,\Msol\ and Kroupa IMF (named RB-SSP-Kroup-MU100 models). These models fall both in quadrants II (WR Galaxies/BCD Galaxies) and III (SB Galaxies/HII Galaxies).

To constrain our models we compare our results with the optical, MIR and FIR observations
available for a few \hii\ galaxies, SB and AGN (Table~\ref{tab:observations}), and the optical emission lines of the BAT-AGN sample. We also compare our predictions with spatially resolved {\it JWST}/MIRI observations of two nearby AGNs. 

 We derive the lower boundary limiting models with stellar age from 100 to 1000 Myr and  $0.0002<\zstar<0.06$. We find that low metallicity models lie close to the boundary in quadrants II and III. In particular, most models in quadrant III have the lowest $\zstar$ ($\log\,\zstar\,<\,-3$). We derive a second lower boundary by ignoring the lower $\zstar$ models, which clearly excludes the weak AGNs from the SINGS sample, as well as all the \hii\ galaxies. We conclude that excitation by AGN-driven shocks is most likely the dominant ionising source in some galaxies in quadrant III. We highlight that both the MIR diagnostic diagram and the spatially resolved {\it JWST}/MIRI data are powerful tools to disentangle the different sources of ionization, especially in galaxies with a complex evolutionary dynamic, in which AGNs, shocks, and a mixture of stellar populations may coexist.

In particular, our models reproduce the observed MIR line ratios of NGC\,4569
(Fig.~\ref{fig:mir}), one of the four \hii\ galaxies in the SINGS sample in Table~\ref{tab:observations}, suggesting that mainly low metallicity HOLMES are responsible for the ionization in this galaxy. In Fig.~\ref{fig:mir}, NGC\,5713 and NGC\,4254 are in agreement with shock models, whereas NGC\,3351 lies below the SF boundary and close to the shock boundary. The emission of \oivw\ in NGC\,3351 could point to the presence of a weak AGN, as in the case of NGC\,4536 and NGC\,253, which show evidence of a weak AGN and have similar MIR emission line ratios to NGC\,3351. Only five of the ten \hii\ galaxies with MIR and FIR observations are reproduced by this set of SSP models (Fig.~\ref{fig:fir}). For the remaining objects, the high emission of \oivw\ can be attributed again to the presence of a weak AGN.

From the boundary traced in the BPT diagram by our Kroupa IMF ($M_{u}$\,=\,100\,\Msol) SSP models (Fig.~\ref{fig:regions}),
we derive a new optical demarcation line, which represents the maximum contribution to line emission by massive stars in SB and HOLMES in retired galaxies, without invading the AGN-dominated region. This line predicts a larger contribution to the emission due to the inclusion of the HOLMES far-UV continuum. If this far-UV emission is excluded, our models reproduce the classical demarcation line previously derived by \citet[][]{Kewley01}.

Considering that the far-UV ionising radiation from HOLMES results in emission line ratios that mimic those in AGN, a non-negligible population of (weak) AGNs may be misclassified. Detailed observational studies about the role of this population in the optical and MIR spectral ranges need to be done. MIR spectra from {\it JWST}/MIRI will allow detailed studies of \oivw\ emission in local AGN, SB, and \hii\ galaxies.

\section{Data Availability}
The data underlying this article are available in [the Mexican Million Model database] at [\url{https://sites.google.com/site/mexicanmillionmodels/the-different-projects/cb_19?pli=1}], and can be accessed with the reference code [CB\_19]. 

\section*{Acknowledgements}
Calculations were performed with version 17.03 of {\sc Cloudy}, described by Ferland et\,al. (2017).
This work is based partially on observations made with the NASA/ESA/CSA James Webb Space Telescope. The data were obtained from the Mikulski Archive for Space Telescopes at the Space Telescope Science Institute, which is operated by the Association of Universities for Research in Astronomy, Inc., under NASA contract NAS 5-03127 for JWST. These observations are associated with programs No.\,1039 and 2732.
We thank S. Charlot for useful discussions. GB acknowledges financial support from the National Autonomous University of M\'exico (UNAM) through grants DGAPA/PAPIIT IG100319 and BG100622. CM acknowledges the support from UNAM/DGAPA/PAPIIT grant IN101220. MK acknowledges the support from the Korea Astronomy and Space Science Institute under 
the R\&D program(Project No. 2022184009) supervised by the Ministry of Science and ICT.







\appendix
\section{BPT diagrams}\label{AppxA}
\begin{figure*}
  \centering
  \includegraphics[width=0.825\textwidth]{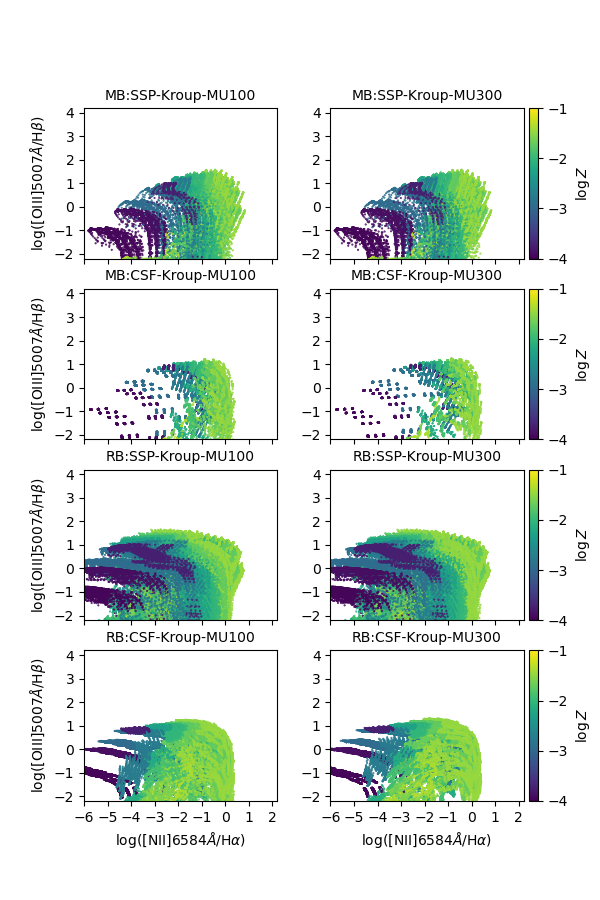}
  \caption{Same as Fig.~\ref{fig:kroupa_mir} but for the optical ratios.}
  \label{fig:kroupa_opt}
\end{figure*}

\begin{figure*}
  \centering
  \includegraphics[width=0.825\textwidth]{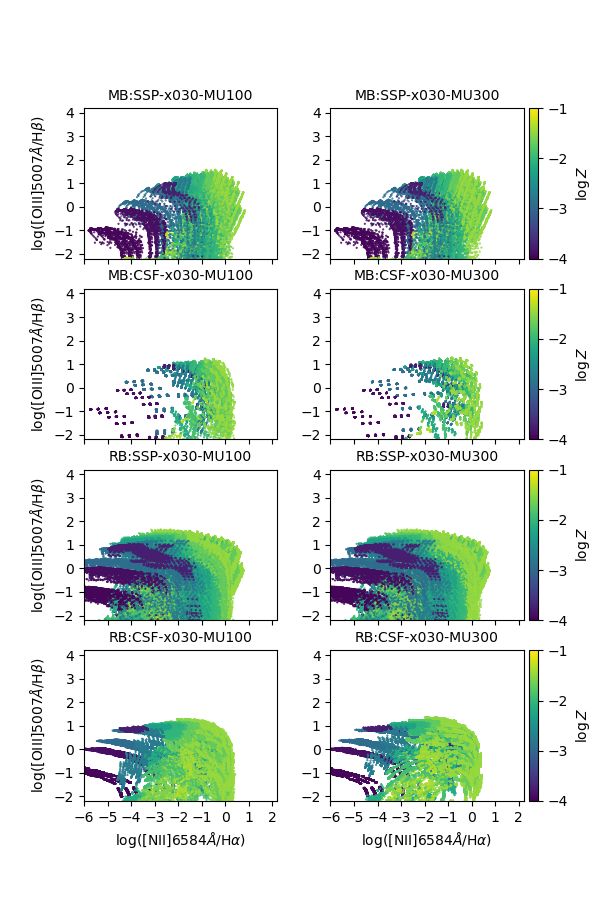}
  \caption{Same as Fig.~\ref{fig:x030_mir} but for the optical ratios.}
  \label{fig:x030_opt}
\end{figure*}

In Figs.~\ref{fig:kroupa_opt} and \ref{fig:x030_opt} we show, respectively, the BPT diagrams for the Kroupa and X030 IMF SSP and CSF models, for both $m_u$\,=\,100 and 300 $\Msol$. Models reaching the MB condition are shown in the {\it top panels}. The {\it bottom panels} show the RB models.

\bsp	
\label{lastpage}
\end{document}